\documentclass[12pt]{article}

\usepackage{amsfonts}
\usepackage{amssymb}
\usepackage{amsmath}

\usepackage{amsthm}

\usepackage{authblk}

\addcontentsline{toc}{section}{Appendices}

\newcommand{\bm}[1]{\mbox{\boldmath $#1$}}

\def\lie{{\pounds}}

\def\S{\Sigma}

\def\be{\begin{equation}}
\def\ee{\end{equation}}
\def\bea{\begin{eqnarray}}
\def\eea{\end{eqnarray}}
\def\bean{\begin{eqnarray*}}
\def\eean{\end{eqnarray*}}

\def\={\stackrel{\Sigma}{=}}

\newlength{\cellwidth}
\usepackage{tabularx}
\usepackage{multirow}
\usepackage{array}

\def\nn{(n\cdot n)}
\def\ll{(\ell\cdot \ell)}

\def\hs{\hspace{5mm}}

\def\H{{\mathcal H}}

\newtheorem{remark}{Remark}

\def\defi{:=}

\def \a {\alpha}
\def \b {\beta}
\def \m {\mu}
\def \n {\nu}
\def \r {\rho}
\def \g {\gamma}
\def \l {\lambda}

\def \d {\delta}

\def \N {\nabla}
\def \Nb {\overline{\nabla}}

\def \Rb {\overline{R}}

\def\deltaS{\underline{\delta}^\Sigma}

\def\otheta{\underline{\theta}}
\def\1{\underline{1}}

\def\H{{\cal H}}

\newcounter{mnotecount}[section]

\renewcommand{\themnotecount}{\thesection.\arabic{mnotecount}}
\newcommand{\mnote}[1]
{\protect{\stepcounter{mnotecount}}$^{\mbox{\footnotesize
$
\bullet$\themnotecount}}$ \marginpar{
\raggedright\tiny\em
$\!\!\!\!\!\!\,\bullet$\themnotecount: #1} }

\newcommand{\mnotex}[1]
{\protect{\stepcounter{mnotecount}}$^{\mbox{\footnotesize $\bullet$\themnotecount}}$ 
\marginpar{
\raggedright\tiny\em
$\!\!\!\!\!\!\,\bullet$\themnotecount: #1} }

\begin{document}

\title{Equations for General Shells}
\author{Jos\'e M. M. Senovilla\\
Departamento de F\'isica Te\'orica e Historia de la Ciencia, Universidad del Pa\'is Vasco UPV/EHU, Apartado 644, 48080 Bilbao, Spain}

\maketitle

\vspace{-0.2em}

\begin{abstract}
The complete set of (field) equations for shells of arbitrary, even changing,  causal character are derived in arbitrary dimension. New equations that seem to have never been considered in the literature emerge, even in the traditional cases of everywhere non-null, or everywhere null, shells. In the latter case there arise field equations for some degrees of freedom encoded exclusively in the distributional part of the Weyl tensor. For non-null shells the standard Israel equations are recovered {\em but not only}, the additional relations containing also relevant information. The results are applicable to a widespread literature on {\em domain walls}, {\em branes} and {\em braneworlds}, {\em gravitational layers}, {\em impulsive gravitational waves}, and the like. Moreover, they are of a geometric nature, and thus they can be used in any theory based on a Lorentzian manifold.
\end{abstract}

\section{Introduction}
In gravitation, surface layers or thin shells are important idealized objects designed to describe localized concentration of matter, or energy, on a given hypersurface. They are suitable to describe domain walls, brane-worlds\footnote{In this paper ``brane-world'' or ``brane'' will always refer to hypersurfaces ---co-dimension one membranes--- and thus they correspond to string-theory $p$-branes only for $p+1$ one less than the spacetime dimension. Of course, sometimes $p$-branes for other values of $p$ can be thought of as effectively 1-co-dimensional with just one extra dimension ---that can be large or small depending on fundamental constants--- if the rest of dimensions may be ignored, e.g. \cite{MK}, and these cases could also be included. The equations of motion of $p$-branes are usually derived from world-volume actions generalizing the Nambu-Goto or Polyakov string actions, see e.g. \cite{Ortin}, and how these combine with the field equations derived here would be worth exploring, connecting the supergravity-oriented literature with the gravitational one. However, this is out of the scope of this paper.}, impulsive waves, thin layers of matter or of gravitational fields, concentrated lightlike signals, propagation of null matter, etcetera. Since the pioneering work by Lanczos \cite{La,La1}, their properties have been used in a vast literature, usually following the well-known and fundamental {\em Israel equations} \cite{I} for the case of non-null shells. This was later extended to the study of null shells in \cite{CD,BI}, see \cite{BH,BH1} and references therein.

In \cite{MS}, the questions of the junction conditions and of the existence of thin shells supported on hypersurfaces of {\em arbitrary, even changing}, causal character were addressed. Such hypersurfaces or shells are called {\em general}. The importance of dealing with hypersurfaces with possibly changing signature resides in the fact that they actually appear in many physical situations. Furthermore,  because they are the {\em generic} type of hypersurface in any given spacetime. Some examples of relevant general hypersurfaces are, for instance, the spherically symmetric apparent horizon in Vaidya's spacetime ---if the radiation stops at some times; more generally, dynamical horizons which eventually merge with the event horizon in asymptotically flat black-hole spacetimes; any achronal boundary, that is to say, the boundary of any future set \cite{HE}; Kerr's stationary limit surface, which is timelike everywhere but tangent to the event horizon at two lines intersecting the axis of symmetry, where it is null \cite{HE,Wald}; the interfaces arising in phase transitions if they occur in a concentrated spatial region and then propagate causally; signature changing braneworlds; and there are many others. 

Thus, for instance, if one wished to analyze the possibility of placing a surface layer on the stationary limit surface of a Kerr black hole, this should be done on a general, signature-changing, hypersurface. 
Similarly, descriptions of signature change in the brane scenario ---if we happen to live in a brane of a higher-dimensional world--- are described by general shells, see \cite{MSV1,MSV,MSV2}. The case of achronal boundaries is of particular interest because intuitive arguments lead to the expectation that an initial value problem for the gravitational field equations on such boundaries is well posed, see the interesting discussion in \cite{mars}. 
In addition to the above, a unifying framework for all type of shells, independently of their causal character, is desirable, and this may be achieved by describing general shells.

The purpose of this paper is triple: (i) to find the full set of equations for general shells, (ii) to do it in arbitrary spacetime dimension, and (iii) the completion of known results for constant-signature shells. The fundamental object describing a thin shell (of arbitrary causal character) is the jump of the derivatives {\em transversal} to the shell of the tangential components of the metric \cite{I,CD,BI,MS}. This is encoded on a symmetric tensor field $y_{ab}$ defined only on the shell and tangent to it, and intrinsic to the shell. This tensor field can be put in correspondence with jumps of well-defined geometric objects, such as the second fundamental form (for non-null shells), or the derivative of a certain one-form in the shell (for null shells). A combination of the two provides the general expression for arbitrary shells, given in (\ref{ygen}). This tensor $y_{ab}$ carries the entire information contained in the singular, distributional part, of the curvature tensor distribution (i.e., on the term proportional to a Dirac-delta type distribution with support on the shell).

The field equations for the shell are by definition the equations satisfied by this tensor $y_{ab}$. Given its direct relationship with the distributional part of the curvature, such equations are derived from those satisfied by the curvature itself, and these are simply the second Bianchi identities, which were shown to hold in the distributional sense for general shells in \cite{MS}. The full set of equations satisfied by $y_{ab}$ derived from the Bianchi identities are presented in section \ref{sec:eqs} for general shells. These can be seen to contain both algebraic and differential equations.

A formula expressing the fundamental tensor $y_{ab}$ in terms of the distributional part of the curvature ---and not the other way round---, as a matter of fact in terms of the distributional parts of the Weyl and Einstein tensor distributions, is presented in section \ref{sec:WE} for general shells. Their simplified expressions at null, and non-null, points of the shell are also discussed. At null points there arise a Weyl singular part representing a typical transversal gravitational wave signal. The relations between the Weyl and Einstein distributional singular parts are also obtained fully explicitly for general shells.

Using the expressions of $y_{ab}$ in terms of the singular parts of the Weyl and Einstein tensors, the field equations of the shell can be rewritten in terms of the latter. This is done in section \ref{sec:analysis} for general shells. There are algebraic, evolution and constraint type equations. A deeper analysis is performed for null and non-null shells separately. To that end, I decompose the Weyl tensor in canonical parts associated to the chosen orthonormal, or null, bases, depending on the case. Consequently, the field equations for non-null shells can be expressed in terms of the jumps across the shell of the `electric and magnetic' parts of the Weyl tensor \cite{s-e,E-H,HOW,JSS}. Similarly, for the null case, they can be expressed in terms of the jumps of the boost-weighted Weyl and Einstein components in a given null frame \cite{CMPP,MCPP}, {\em except} for the components with boost weight $-2$, which do not enter in any of the equations. 

The Israel equations \cite{I} and their generalizations to null \cite{BI} and general shells \cite{MS} are re-obtained. However, these are just a small subset of the full set of field equations, and many other equations are revealed. Some of them have been previously treated in the literature in 4 \cite{BH,BH1} or 5 dimensions \cite{MK,SMS}, or in general dimension for null shells \cite{Marstalk} but, as far as I am aware, the full set of equations, in general dimension, has never been considered before. In this sense, results concerning shells, or branes, or domain walls, or impulsive gravitational waves may have to be revisited and/or completed in the light of the full set of equations. Among the full set of field equations some are specially remarkable, in particular the differential and algebraic equations for the transversal gravitational-wave degrees of freedom that arise in null shells (formulas (\ref{[Jsym]}), (\ref{[Jasym]}), (\ref{[Ricci]}) and (\ref{[L]}) below), as well as the differential equation (\ref{[H]}) and the algebraic equations (\ref{[E]}) and ---in dimension higher than 4--- (\ref{[F]}) for non-null shells.

A brief concluding section \ref{Conclusions} closes the paper, where some important remarks are placed. In particular, I stress that the full set of shell field equations herein presented are of pure geometric character, and therefore serves as a theoretical framework valid in any theory based on a Lorantizan manifold. The way to apply them to specific gravitational field equations is succinctly explained with relevant examples. Besides, the proper matching conditions (that is, with no distributional energy-momentum tensor) for general hypersurfaces  are also easily determined from the given equations. 

An appendix with all necessary material from previous works is added for the benefit of the reader.

\section{Basic formulas}\label{sec:basic}
Let $(M,g)$ denote a connected, oriented and time-oriented $(m+1)$-dimensional Lorentzian manifold with metric $g$ of signature $(-,+,\dots,+)$. Lowercase Greek letters $\alpha,\beta,\dots $ are spacetime indices and run from $0$ to $m$. Small Latin indices $a,b,\dots$ are hypersurface indices and take values from $1$ to $m$. 

I assume that $(M,g)$ has been constructed as the junction of two given spacetimes $(M^\pm , g^\pm)$ across a matching hypersurface $\S$, which is the shell, and that the resulting spacetime has a continuous metric $g$ (so that, in appropriate coordinates, $g^+\=g^-$). This requires that the first fundamental forms $\bar{g}^\pm$ inherited by $\S$ from $(M^\pm , g^\pm)$ agree and, in the case that $\S$ is null somewhere, furthermore the existence of transversal vectors fields $\vec{\ell}^\pm$ on each side of $\S$, $\vec\ell^\pm$ having the same scalar products with any given basis of tangent vectors at any $x\in \S$, and such that $\vec{\ell}^+$ points towards $M^+$ while $\vec{\ell}^-$ points outwards from $M^-$ (see \cite{MSV}, which corrects previous partial statements in \cite{CD,MS}).

Denote by $\{\vec{e}_a\}$ a chosen basis of vector fields in $T_\S(M)$, that is, tangent to $\S$ on $\S\subset M$.\footnote{If the embeddings are explicitly given in parametric form in local coordinates $\{x_\pm^\m\}$ on $M^\pm$ and $\{\xi^a\}$ on $\S$, then these vector fields can be chosen as $e^\mu_a = \partial x_\pm^\mu/\partial \xi^a$, and they must agree on both $\pm$ sides. This is very helpful in explicit calculations, but not necessary for this note.}
Notice that $\{\vec{e}_a\}$ are defined only on $\S$. Let $n^\pm_\mu$ be two normal one-forms to $\S$, one for each side. They are fixed up to an overall non-vanishing factor by the conditions
$$
n^\pm_\mu e^\mu_a=0,
$$
and I choose $n^-_\mu$ pointing outwards from $M^-$ and $n^+_\mu$ pointing towards $M^+$. Observe that, for general hypersufaces (containing points where $\S$ is null), these one-forms cannot be normalized: actually, the corresponding ``normal'' vector fields $n_\pm^\mu$ are tangent to $\S$ on those null points. Thus, to complete the tangent bases one must choose {\em rigging} vector fields \cite{Sch,MS} $\ell_\pm^\mu\in T_\S M^\pm$ on $M^\pm$, defined by being transversal to $\S$ everywhere on $\S$. Then, a convenient normalization choice is
$$
n^\pm_\mu \ell_\pm^\mu =1. 
$$
The agreement of the two $(\pm)$-first fundamental forms amounts to the equalities on $\Sigma$
$$
\bar{g}^+_{ab} = \bar{g}_{ab}^- \defi \bar{g}_{ab},\hspace{1cm}
\bar{g}^\pm_{ab} := g_{\mu\nu}^\pm |_\S e^\mu_a e^\nu_b
$$
so that the tangent vector fields $\{\vec{e}_a\}$ have equal scalar products on $\S$ from both sides and $\bar{g}_{ab}$ represents the (unique) first fundamental form on $\S$. The riggings are then fixed by demanding
$$
g_{\mu\nu}|_\S \ell^\mu_+ e^\nu_a = g_{\mu\nu}|_\S \ell^\mu_- e^\nu_a\defi \ell_a, \hspace{1cm}
g_{\mu\nu}|_\S \ell^\mu_+ \ell^\nu_+ = g_{\mu\nu}|_\S \ell^\mu_- \ell^\nu_-\defi \ll
$$
so that the two bases on the tangent spaces 
$$
\{\vec{\ell}^{+},\vec{e}_a\}\equiv  \{\vec{\ell}^{-},\vec{e}_a\}\defi \{\vec{\ell},\vec{e}_a\}
$$
are then identified and the $\pm$ are dropped.\footnote{This identification is usually only abstract, for in practice one still uses bases on $\pm$-coordinate systems to do explicit calculations.} The two one-forms $\bm{n}_\pm$ are also identified so that the dual basis is denoted simply by
$$
\{n_\mu, \omega^a_\mu\}
$$
where the one-forms $\bm{\omega}^a$, which depend on the choice of rigging, are characterized by
$$
\ell^\mu \omega^a_\mu =0, \hspace{1cm} e^\mu_b \omega^a_\mu =\delta^a_b .
$$

I am going to closely follow the notation in \cite{MS}, in particular I will use the abbreviation
$$
\nn \defi n_\mu n^\mu 
$$
and square brackets for the difference (the {\em jump}) of any object as computed from the $+$ or the $-$ sides of $\S$, namely
\be
\left[T\right] (p) \defi \lim_{x\stackrel{M^+}{\rightarrow} p} T^+(x)- \lim_{x\stackrel{M^-}{\rightarrow} p} T^-(x), \hspace{1cm} \forall p\in \S \label{discont}
\ee
where $T^\pm$ represent either the values of $T$ at $M^\pm$ if they exist there, or the versions of $T$ as inherited on $\S$ from $M^\pm$ if they are defined only on $\S$, respectively. Similarly, under the same hypotheses, the value of $T$ on $\S$ is denoted by
\be
T^\S (p):= \frac{1}{2} \left( \lim_{x\stackrel{M^+}{\rightarrow} p} T^+(x)+ \lim_{x\stackrel{M^-}{\rightarrow} p} T^-(x)\right), \hspace{1cm} \forall p\in \S .\label{TSigma}
\ee

A list of the main formulas to be used or needed in what follows are collected in the Appendix. Readers not familiar with the spacetime matching procedure may find it useful to read the appendix before continuing reading the main text. I will say that $\S$ is a {\em standard} hypersurface if the curvature tensor is at most discontinuous there. The basic objects measuring the departure of $\S$ from a standard hypersurface are given by $\H^\pm_{ab}$ as defined in (\ref{H}). These depend on the choice of the rigging, and their anty-symmetric parts do not jump on $\S$: $[\H_{[ab]}]=0$. Its symmetric part does, providing the fundamental symmetric tensor defining the properties of the shell, given in (\ref{y}):
\be
y_{ab} \defi \left[ \H_{(ab)}\right] . \tag{\ref{y}}
\ee
It is very important to remark that $y_{ab}$ does {\em not} depend on the choice of rigging \cite{MS}.
This tensor, or equivalently its spacetime version $y_{\m\n}=\omega^a_\m \omega^b_\n y_{ab}$, encodes all the information contained in the shell, because the singular ``Dirac-delta''-type part of the Riemann tensor distribution (\ref{Riedist}) is fully determined by $y_{\m\n}$, see (\ref{Hriemann}). In particular, $y_{ab}$ vanishes if and only if the Riemann tensor is a (maybe discontinuous) tensor field, that is to say, if and only if $\S$ is a standard hypersurface.

\subsection{How to recover the shells with constant signature}
If one wishes to specialize to the standard cases where the first fundamental form has constant signature everywhere on $\S$, one simply has to restrict the formulas according to the next rules.
\subsubsection{Non-null shells}
For everywhere timelike or spacelike shells $\S$ there is a canonical choice of rigging given by the normal vector field itself. Then, simply set (with $\epsilon \in \{-1,1\}$)
\bea
\bm{\ell} =\epsilon \bm{n}, \hs \nn =\epsilon =\ll ,\hs  \ell_a =0 =n^a, \hs \varphi^\pm_a=0, \label{nonnull1}\\
\H^\pm_{ba} =\H^\pm_{(ab)}= \bar{g}_{ac} \Psi^c_{\pm b} = \epsilon K^\pm_{ab}, \hs y_{ab} =\epsilon \left[ K_{ab}\right].\label{nonnull2}
\eea
where $\epsilon =1$ corresponds to timelike $\S$ and $\epsilon =-1$ to a spacelike one. In these cases, the rigged connection $\bar{\Gamma}$ coincides with the Levi-Civita connection of $\bar{g}_{ab}$ which defines also a canonical volume element so that there is a preferred decomposition (\ref{delta}) intrinsically defined. 
\subsubsection{Null shells}
If the shell $\S$ is null everywhere, then one needs to adapt the formulas by setting
\bea
\nn =0, \hs n^\a = n^a e^\a_a, \hs \bar{g}_{ab}n^b =K_{ab}n^b =0, \hs \ell_a\bar{g}^{ab} =-\ll n^b,\label{null1}\\
(\lie_{\vec n} \bar{g})_{ab} =2 K_{ab}, \hs \left[K_{ab} \right]=0, \hs y_{ab} =\left[ \Nb_{(a} \ell_{b)}\right].\label{null2}
\eea
Observe that in this case one can also choose the rigging $\vec\ell$ to simplify further some expressions, for instance by setting $\ll =0$ or $\ll =\pm 1$. A typical and useful well-adapted choice is that of a null rigging $\ll =0$; this will be called a {\em null gauge}, and one should keep in mind that it is not unique as null rotations leaving $\vec n$ invariant are permitted. For the time being, however, I do not choose a null gauge and I keep the choice of rigging $\vec\ell$ free.

\section{Field equations for general shells}\label{sec:eqs}
In this paper, the field equations for a shell are by definition the differential equations satisfied by the singular part of the curvature tensor distribution, or equivalently, the differential equations satisfied by the fundamental tensor $y_{ab}$. These lead in particular, via the gravitational field equations (I am not assuming General Relativity or any other particular theory), to the equations for the {\em singular} part of the energy-momentum tensor distribution with support on $\S$ (for instance, in General Relativity these are simply given by the singular part of the Einstein tensor distribution divided by the gravitational constant). 

There are three possible routes to obtain the field equations obeyed by shells in spacetimes: (i) by resorting to the use of tensor distributions, (ii) by comparing the Gauss and Codazzi equations from both sides of $\S$, or (iii) by adding proper boundary terms on $\S$ to the action. I am going to present the first two ways in full in what follows, to show that they fully agree and for completeness. However, I will not consider the third route ---despite its popularity in the braneworld literature--- for a number of reasons. Firstly, my approach is purely geometric and therefore independent of any field equations, hence the action of the theory is unknown and left free. Secondly, even if one had the action available, there are problems in finding the correct boundary terms for different theories, including some basic ambiguities \cite{D,GCT,NO,BD,BD1}; furthermore, it is customary to use Gaussian coordinates based on the brane which is unfortunately not appropriate for a derivation of the correct field equations as explained in the Appendix of \cite{RSV}, missing for instance the gravitational double layers that arise in quadratic theories of gravity \cite{Senovilla13,S2,S3,RSV}. Thirdly, because the problem of the boundary terms is much more difficult for the case of null shells \cite{Ch}, not to speak of {\em general shells}: only very recently there has been some consideration of this question in pure General Relativity showing important ambiguities \cite{LMPS}. In summary, the third route looks intractable and not advisable. I would like to add, however, that the results in this paper may serve as inspiration in the search of the proper boundary terms for general or null boundaries, in generic theories based on a Lorentzian manifold.

\subsection{The field equations derived from a distributional computation}

The general differential conditions on the curvature tensor distribution are the second Bianchi identities 
\be
\nabla_{\lambda}\underline{R}^\a{}_{\b\m\n} +\nabla_{\m}\underline{R}^\a{}_{\b\n\l}+\nabla_{\n}\underline{R}^\a{}_{\b\l\m}=0 \label{bianchi}
\ee
which were proven to hold in full generality for spacetimes with a metric just continuous across $\S$ in \cite{MS}. Using here (\ref{Riedist}), (\ref{deltamu}) and (\ref{nablaT1}) one readily gets
\be
\nabla_{\lambda}\underline{H}^\a{}_{\b\m\n} +\nabla_{\m}\underline{H}^\a{}_{\b\n\l}+\nabla_{\n}\underline{H}^\a{}_{\b\l\m}=-\left[R^\a{}_{\b\m\n}\right] \underline\delta_{\l} - \left[R^\a{}_{\b\n\l}\right] \underline\delta_{\m}-\left[R^\a{}_{\b\l\m}\right] \underline\delta_{\n} \label{FE}
\ee
where $\underline{H}^\a{}_{\b\m\n}$ is defined in (\ref{Hriemann}). This can also be expressed as
\be
\nabla_{\lambda}\left(H^\a{}_{\b\m\n} \deltaS\right)+\nabla_{\m}\left(H^\a{}_{\b\n\l}\deltaS\right)+\nabla_{\n}\left(H^\a{}_{\b\l\m}\deltaS\right)=-\deltaS\left(\left[R^\a{}_{\b\m\n}\right] n_{\l} + \left[R^\a{}_{\b\n\l}\right] n_{\m}+\left[R^\a{}_{\b\l\m}\right] n_{\n} \right).\label{FE1}
\ee
Either (\ref{FE}) or (\ref{FE1}) contain all the information concerning the field equations of the shell. To extract this information a standard calculation in tensor distributions allows one to prove (see, e.g., appendix D.3 in \cite{RSV} for this calculation in the case of non-null $\S$)
$$
\nabla_{\lambda}\left(H^\a{}_{\b\m\n} \deltaS\right)= \nabla_\rho \left(H^\a{}_{\b\m\n}n_\l \ell^\rho \deltaS \right)+\deltaS \left(P_\l^\rho \nabla^\S_\rho H^\a{}_{\b\m\n}-\Psi^c_{\S c} H^\a{}_{\b\m\n}n_\l +\varphi^\S_c \omega^c_\l H^\a{}_{\b\m\n} \right)
$$
so that inserting this into (\ref{FE1}) and using (\ref{Hn}) one arrives at
\bea
-\left[R^\a{}_{\b\m\n}\right] n_{\l} - \left[R^\a{}_{\b\n\l}\right] n_{\m}-\left[R^\a{}_{\b\l\m}\right] n_{\n}&=&
P_\l^\rho \nabla^\S_\rho H^\a{}_{\b\m\n}+P_\m^\rho \nabla^\S_\rho H^\a{}_{\b\n\l}+P_\n^\rho \nabla^\S_\rho H^\a{}_{\b\l\m}\nonumber\\
&+& \varphi^\S_c \left(\omega^c_\l H^\a{}_{\b\m\n} +\omega^c_\m H^\a{}_{\b\n\l}+\omega^c_\n H^\a{}_{\b\l\m}\right).\label{FE2}
\eea
These constitute the full set of field equations on the shell $\S$, relating the derivatives of the $\S$-singular part of the curvature tensor distribution to the jumps of the curvature tensor across $\S$. Both sides of these equations depend on how one chooses the normalization factor for $\bm{n}$ but they are affected in exactly the same manner and thus the field equations hold for every such choice. Observe that the lefthand side is independent of the choice of rigging, and thus so is the righthand side ---despite the single terms there depending on such choice.

By contracting (\ref{FE2}) with $\ell^\l e^\m_a e^\nu_b$ one gets the following alternative, equivalent, form of the field equations
\be
-\left[R^\a{}_{\b\m\n}\right] e^\m_a e^\nu_b= \ell^\n e^\m_be^\sigma_a\nabla^\S_\sigma H^\a{}_{\b\m\n} -
\ell^\n e^\m_ae^\sigma_b\nabla^\S_\sigma H^\a{}_{\b\m\n}+\ell^\n H^\a{}_{\b\m\n}\left(\varphi^\S_a e^\m_b -\varphi^\S_b e^\m_a \right). \label{FE3}
\ee
In this form, the field equations are independent of the choices of normal one-form $\bm{n}$ and of the rigging $\vec\ell$.

Either (\ref{FE2}) or (\ref{FE3}) contain $(m+4)(m+1)m(m-1)/12$ independent relations (14 in dimension $m+1=4$). These are differential equations for the tensor field on $\S$ defined in (\ref{y}), which itself contains $m(m+1)/2$ independent components at most. To write the equations in terms of $y_{ab}$ explicitly, contract (\ref{FE3}) with $\omega^d_\a e^\b_c$, $\omega^c_\a\ell^\b$ and $n_\a e^\b_c$ to obtain, respectively
\bea
-\omega^d_\a e^\b_c \left[R^\a{}_{\b\m\n}\right] e^\m_a e^\nu_b=\overline\nabla^\S_a (n^d y_{cb})-\overline\nabla^\S_b (n^d y_{ca})+\nn\left(\Psi^d_{\S a} y_{cb} -\Psi^d_{\S b} y_{ca}  \right)+K^\S_{cb} y^d_a -K^\S_{ca} y^d_b,\label{eq1}\\
\omega^c_\a\ell^\b\left[R^\a{}_{\b\m\n}\right] e^\m_a e^\nu_b=2\overline\nabla^\S_{[a} y^c_{b]}+2y^c_{[a}\varphi^\S_{b]}+2 \Psi^c_{\S [a} y_{b]d}n^d+2 n^c \Psi^d_{\S[a}y_{b]d}, \hspace{3cm} \label{eq2}\\
-n_\a e^\b_c \left[R^\a{}_{\b\m\n}\right] e^\m_a e^\nu_b=2\overline\nabla^\S_{[a}\left(\nn y_{b]c} \right)+2n^dy_{d[a} K^\S_{b]c}-2n^dK^\S_{d[a}y_{b]c}+2\nn \varphi^\S_{[a} y_{b]c}.\hs  \label{eq3}
\eea
Here 
\be
y^c_b \defi \omega^c_\mu y^\mu_{\nu} e^\nu_b = \omega^c_\mu g^{\mu\rho} y_{\rho\nu} e^\nu_b=
\left(n^c \ell^\rho + \bar{g}^{cd}e^\rho_d \right)y_{\rho\nu} e^\nu_b=\bar{g}^{cd} y_{db}.\label{yup}
\ee
A further contraction with $n_\a \ell^\b$ leads to another expression 
\be
n_\a \ell^\b \left[R^\a{}_{\b\m\n}\right] e^\m_a e^\nu_b=\overline\nabla^\S_a (n^c y_{cb})-\overline\nabla^\S_b (n^c y_{ca})+\nn\left(\Psi^c_{\S a} y_{cb} -\Psi^c_{\S b} y_{ca}  \right)+K^\S_{cb} y^c_a -K^\S_{ca} y^c_b
\label{eq4}
\ee
not independent of (\ref{eq1}) because the contraction of $c$ and $d$ there leads to (\ref{eq4}) through (\ref{P}). 

\subsection{The field equations derived from the Gauss-Codazzi relations}
One can also find the shell field equations by using the Gauss and Codazzi relations, as done originally by Israel for non-null hypersurfaces in \cite{I}. Start, for instance, from (\ref{gauss}) and subtract the $\pm$ lefthand sides to get
$$
\omega^d_\a \left[R^{\a}{}_{\b \g \d}\right] e_a^\b e_b^\g e_c^\d = \left[\Rb^{d}{}_{abc}\right] - K^\S_{ac}\left[\Psi_{b}^d\right] -\left[K_{ac}\right] \Psi^d_{\S b}+ K^\S_{ab}\left[\Psi^d_{c}\right]+\left[K_{ab}\right]\Psi^d_{\S c}
$$
and, for the first summand on the righthand side, use the last in (\ref{jumps}) to find
$$
\left[\Rb^{d}{}_{abc}\right] =\overline\nabla^\S_b\left(n^d y_{ca} \right)-\overline\nabla^\S_a\left(n^d y_{cb} \right)
$$
and inserting this together with the first and third in (\ref{jumps}) into the previous formula, and using (\ref{yup}), one readily arrives at exactly (\ref{eq1}). 

Starting now from (\ref{codazzi2}), subtracting the lefthand sides and noting that 
$$
\left[ \overline\nabla_a \Psi^c_b\right]= \overline\nabla^\S_a \left[\Psi^c_b\right]+n^d y_{ab} \Psi^c_{\S d}-n^c y_{ad}\Psi^d_{\S b}
$$
one easily gets (\ref{eq2}) by using the first and second in (\ref{jumps}). 

Similar simple calculations show that the subtraction of the lefthand sides in (\ref{codazzi1}) and (\ref{codazzi3}), lead, respectively, to (\ref{eq3}) and (\ref{eq4}). 

\subsection{Remarks and alternative forms of the field equations} 
Some remarks are in order:
\begin{enumerate}
\item the covariant derivatives in (\ref{eq1}) are not present in the case of non-null $\S$ if the choice of canonical gauge is made, providing a direct relation between the jumps of the Riemann tensor tangential part with $y_{ab}=\epsilon \left[ K_{ab}\right]$, while (\ref{eq2}) gives a relation between partly tangential but partly orthogonal Riemann tensor jumps and the derivatives of $y_{ab}$. Equation (\ref{eq3}) is then simply redundant.
\item In the case of everywhere null $\S$, the covariant derivatives are not present in (\ref{eq3}), providing a direct relation of $n^by_{ba}$ with the jump $n_\a e^\b_c \left[R^\a{}_{\b\m\n}\right] e^\m_a e^\nu_b$. Observe that such a jump can also be given as that of purely tangential components of the curvature tensor with all indices down in this case:
$$
n_\a e^\b_c \left[R^\a{}_{\b\m\n}\right] e^\m_a e^\nu_b=n^\a e^\b_c \left[R_{\a\b\m\n}\right] e^\m_a e^\nu_b=
n^a e^\a_a e^\b_c \left[R_{\a\b\m\n}\right] e^\m_a e^\nu_b\, .
$$
The remaining components of $y_{ab}$ not included in $y_{ab}n^b$ are related to curvature jumps via differentiation as well, as in (\ref{eq2}). Again (\ref{eq3}) can be seen to be redundant, being a linear combination of (\ref{eq2}) and (\ref{eq1}) using (\ref{omegaup}). 
\item For general hypersurfaces with changing causal character it is better to keep the three expressions, as there is no canonical global choice of rigging and, moreover, there are situations in which the closed set defined by $\{s\in \S, \nn |_s=0\}$ may contain parts with empty interior, and thus even though $\nn=0$ there, its derivative is non-zero. 
\end{enumerate}
In summary, (\ref{eq2}) and (\ref{eq1}) can be taken as the fundamental set of equations in all possible cases. There are several interesting and more useful ways to write down these two fundamental equations. For instance, inserting (\ref{yup}) into (\ref{eq2}) and using (\ref{nablagup}) one easily finds
\be
\omega^c_\a\ell^\b\left[R^\a{}_{\b\m\n}\right] e^\m_a e^\nu_b=\bar{g}^{cd}\left(\overline\nabla^\S_a y_{bd}-\overline\nabla^\S_b y_{ad} +y_{da}\varphi^\S_b - y_{db}\varphi^\S_a\right). \label{eq2'}
\ee
Multiplying here by $\bar{g}_{ce}$, using (\ref{gup}) and (\ref{omegaup}) and noting that the terms that arise proportional to $\ell_e$ cancel due to (\ref{eq4}) and (\ref{nablan}), this can still be rewritten as
\be
-\ell^\a \left[ R_{\a\b\m\n}\right]e^\b_c e^\m_a e^\n_b =\overline\nabla^\S_a y_{bc} -\overline\nabla^\S_b y_{ac} +y_{ca}\varphi^\S_b -y_{cb} \varphi^\S_a. \label{eq2''}
\ee
Concerning (\ref{eq1}), on using (\ref{nablan}) on the righthand side it becomes
$$
-\omega^d_\a e^\b_c \left[R^\a{}_{\b\m\n}\right] e^\m_a e^\nu_b=2n^d\left(\overline\nabla^\S_{[a} y_{b]c}+y_{c[a}\varphi^\S_{b]} \right)+
\bar{g}^{de} \left(K^\S_{ea}y_{cb}-K^\S_{eb}y_{ca}+y_{ea}K^\S_{cb}-y_{eb}K^\S_{ca}\right)
$$
and using here (\ref{eq2''}) together with (\ref{omegaup}) one obtains
\be
\bar{g}^{de} \left[ R_{\a\b\m\n}\right]e^\a_d e^\b_c e^\m_a e^\n_b=
\bar{g}^{de} \left(K^\S_{db}y_{ca}-K^\S_{da}y_{cb}-y_{da}K^\S_{cb}+y_{db}K^\S_{ca}\right).\label{eq1'}
\ee
Multiplying here by $\bar{g}_{ef}$, using (\ref{na}), (\ref{eq3}), (\ref{eq2''}), (\ref{nablann}) and after a straightforward calculation this can be rewritten as
\be
\left[ R_{\a\b\m\n}\right]e^\a_d e^\b_c e^\m_a e^\n_b=K^\S_{db}y_{ca}-K^\S_{da}y_{cb}-y_{da}K^\S_{cb}+y_{db}K^\S_{ca}. \label{eq1''}
\ee

Eqs.(\ref{eq2''}) and (\ref{eq1''}) are valid for {\em general} shells. The former could also be easily derived starting from equation (17) in \cite{MS} for both $\pm$-sides, substracting them and following a procedure analogous to the one used in the previous subsection for the Gauss and Codazzi equations. Eqs.(\ref{eq1}-\ref{eq3}), as well as (\ref{eq2''}), are written in terms of the mean rigged connection $\overline{\Gamma}^\S$ and corresponding covariant derivative $\Nb^\S$ as defined in the appendix. However, $\Nb^\S$ is by no means a unique or preferred covariant derivative and any other connection well defined on $\S$ could be used. For instance, a ``rigged metric connection'' ---associated to any non-null rigging--- introduced in \cite{MS} is a viable possibility. In fact, such connection has the advantage that it has no jump across $\S$. Observe, in this sense, that the used rigged covariant derivative $\Nb^\S$ is built from the rigged connections inherited by $\S$ from both sides $M^\pm$, and in this sense it contains information included in the fundamental tensors $Y^\pm_{ab}=\H^\pm_{(ab)}$ defined in (\ref{Y}) ---and whose jump is $y_{ab}$. This suggests that a particularly suited choice of connection is given by the ``metric hypersurface connection'' defined in \cite{mars}, whose main feature is precisely that is fully independent of $Y^\pm_{ab}$. Actually, expressions equivalent to (\ref{eq2''}) and (\ref{eq1''}) follow easily, in terms of such a metric hypersurface connection, from Proposition 6 in \cite{mars} by simply putting square brackets on both sides.

Eqs.(\ref{eq1}-\ref{eq3}), as well as (\ref{eq2''}) and (\ref{eq1''}), can also be rewritten in terms of the {\em singular parts} of the Einstein and Weyl tensor, or other tensors derived from the curvature. To find these forms, one needs to write down the relation between $y_{ab}$ and the singular parts of the Weyl and Einstein tensors. This is done in the next section, while a deeper analysis of the field equations is left for section \ref{sec:analysis}.

\section{The Weyl and Einstein tensor distributions}\label{sec:WE}
As announced above, we need the relation between $y_{ab}$ and the singular parts of the Weyl and Einstein tensors. 
\subsection{The Einstein tensor distribution singular part}
With regard to the the Einstein tensor distribution, by taking the part of (\ref{G}) projected to $\S$ one gets
\be
{\cal G}_{ab} \defi {\cal G}_{\m\n}e^\m_a e^\n_b= -\nn y_{ab} -\overline{g}_{ab} \left(n^c n^d -\nn \bar{g}^{cd} \right) y_{cd}.\label{G1}
\ee
Analogously, a straightforward calculation leads to
\bea
{\cal G}^a{}_b \defi \omega^a_\a {\cal G}^\a_\b e^\b_b= \left(n^c n^d -\nn \bar{g}^{cd} \right)\left(\delta^a_c y_{bd} -\delta^a_b y_{cd} \right), \hspace{2cm} \label{G2}\\
{\cal G}^{ab}\defi \omega^a_\a \omega^b_\b {\cal G}^{\a\b} =-\nn y^{ab} +n^a y^b_cn^c +n^b y^a_c n^c +y^c_c \left(\nn \bar{g}^{ab} - n^a n^b \right)-\bar{g}^{ab} n^d n^c y_{dc} \label{G3}
\eea
and the trace of (\ref{G2}) is ($H$ is defined in (\ref{Hscalar}))
\be
{\cal G}^b{}_b =(1-m) \left(n^c n^d -\nn \bar{g}^{cd} \right) y_{cd}=\frac{1-m}{2} H \label{trG}
\ee
so that (\ref{G1}) can be rewritten as
\be
-\nn y_{ab} = {\cal G}_{ab} +\frac{1}{1-m}\bar{g}_{ab}{\cal G}^b{}_b 
\label{yG}
\ee
This informs us that:
\begin{itemize}
\item at {\em non-null} points on $\S$, the basic tensor $y_{ab}$ can be put in direct correspondence with the tangential distributional singular part of the Einstein tensor distribution, both of them having the same information {\em there}. 
\item However, at {\em null} points of $\S$, such tangential part ${\cal G}_{ab}$ of the singular Einstein tensor distribution is actually simply proportional to the degenerate first fundamental form $\bar{g}_{ab}$, and carries {\em no information} concerning the basic tensor field $y_{ab}$ ---apart from its component $n^cn^dy_{cd}$. Still, its ``contravariant'' version ${\cal G}^{\a\b}$, which is tangent to $\S$ (see (\ref{nG})) and fully encoded in ${\cal G}^{ab}$, contains the information relative to $n^by_{ab}$ and $y^c_c$.
\end{itemize}

Alternatively, consider transversal components of the Einstein tensor distribution. From (\ref{G}) its singular part satisfies
$$
\ell^\b {\cal G}_{\b\m}= n^ay_{ab}\omega^b_\m -\bar{g}^{ab}y_{ab} n_\m -\frac{1}{2} H \ell_\m ,
$$
from where 
\bea
\ell^\b {\cal G}_{\b\m}e^\m_b &=& n^ay_{ab} -\ell_b \frac{H}{2}=\ell_a {\cal G}^a{}_b, \label{ellGe}\\
\ell^\b {\cal G}_{\b\m}\ell^\m &=& -\bar{g}^{ab}y_{ab} -\ll \frac{H}{2}= -\ll n^an^b y_{ab}- n^c\ell_c \bar{g}^{ab}y_{ab} =\ell_a \ell_b {\cal G}^{ab},\label{ellellG}\\
\ell^\b {\cal G}_\b^\m \omega_\m^a &=&  n^cy_c^a -n^a \bar{g}^{cb}y_{cb}=\ell_b {\cal G}^{ba} \label{ellGomega} \, .
\eea
Thus: 
\begin{itemize} 
\item at {\em non-null} points of $\S$ these transversal parts simply vanish in the canonical gauge. This fact, in combination with the above, implies that all the information carried by $y_{ab}=\epsilon \left[K_{ab}\right]$ is contained, at non-null points of $\S$, in ${\cal G}_{ab}=-\left[K_{ab}\right] +\overline{g}_{ab} \bar{g}^{cd}\left[ K_{cd}\right]$. 
\item at {\em null} points in $\S$ these transversal parts provide the information contained in $n^ay_{ab}$ plus $y^c_c =\bar{g}^{ab}y_{ab}$. Observe that these are exactly the same parts contained in the contravariant (and tangential) form ${\cal G}^{\m\n}={\cal G}^{ab} e^\m_a e^\n_b$. These are $m+1$ components out of the total $m(m+1)/2$ independent components contained in $y_{ab}$. Hence, there remain $m(m+1)/2-m-1= (m+1)(m-2)/2$ independent components not controlled by the singular part of the Einstein tensor distribution at null points of $\S$.
\end{itemize}

\subsection{The Weyl tensor distribution singular part}
Let us pass to consider the rest of the curvature, encoded in the conformally invariant Weyl tensor. As a tensor distribution, this will have the structure
\be
\underline{C}^\a{}_{\b\m\n} = C^{+\a}{}_{\b\m\n} \otheta +C^{-\a}{}_{\b\m\n} (\1-\otheta) +\underline{W}^\a{}_{\b\m\n}, \label{weyldist}
\ee
where $\underline{W}^\a{}_{\b\m\n}$ is the singular part of the Weyl tensor distribution with support on $\S$ and possesses exactly the same symmetries as a Weyl tensor, in particular, it is traceless: $\underline{W}^\a{}_{\b\a\n}=0$. By using the expression relating the Riemann, Weyl and Einstein tensors
$$
R^\a{}_{\b\m\n} = C^\a{}_{\b\m\n} +\frac{1}{m-1} \left(G^\a_\m g_{\b\n}-G^\a_\n g_{\b\m} -\delta^\a_\n G_{\b\m}+ \delta^\a_\m G_{\b\n}\right)+\frac{R}{m}\left(\delta^\a_\m g_{\b\n}-\delta^\a_\n g_{\b\m}\right)
$$
(where, of course, $(1-m)R/2=G^\rho_\rho$) together with (\ref{Hriemann}), an explicit expression is found
\be
\underline{W}^\a{}_{\b\m\n}=\deltaS \left(2n^\a y_{\b[\m}n_{\n]} +2n_\b  y^\a_{[\n} n_{\m]}
-\frac{2}{m-1}  \left({\cal G}^\a_{[\m} g_{\n]\b}+ \delta^\a_{[\m} {\cal G}_{\n]\b}\right)- \frac{2H}{m} \delta^\a_{[\m} g_{\n]\b}\right)
\defi \deltaS W^\a{}_{\b\m\n} \label{W}
\ee
which relates $W^\a{}_{\b\m\n}$, ${\cal G}_{\a\b}$ and $y_{\a\b}$. Now, the identity (\ref{Hn}) becomes
\be
W^\a{}_{\b[\m\n}n_{\l]}= -\frac{2}{m-1} \left(n_{[\l}{\cal G}^\a_{\m} g_{\n]\b}+ n_{[\l}\delta^\a_{\m} {\cal G}_{\n]\b}\right)- \frac{2H}{m} n_{[\l}\delta^\a_{\m} g_{\n]\b} \label{Wn}
\ee
and contraction of (\ref{W}) with $n_\a$ leads to
\be
n_\a W^\a{}_{\b\m\n}=2\nn y_{\b[\m}n_{\n]} +2n_\b n_\a y^\a_{[\n} n_{\m]} +\frac{2}{m-1} {\cal G}_{\b[\m}n_{\n]} +2\frac{H}{m} g_{\b[\m}n_{\n]} \label{nW}.
\ee
The combination of the last two equations provides immediately 
\be
n_\a W^\a{}_{\b[\m\n}n_{\l]}=0 . \label{nWn}
\ee
This important relation will be analyzed later for the cases of everywhere non-null, and everywhere null, shells. Observe that this is equivalent to
$$
n_\a W^\a{}_{\b\m\n}e^\m_a e^\n_b=0.
$$
There are $m(m+2)(m-2)/3$ independent relations here (or equivalently in (\ref{nWn})) out of the possible maximum total of $(m+3)(m+2)(m+1)(m-2)/12$ independent components of a Weyl tensor.

Contracting (\ref{nW}) with $e^\m_a \ell^\n$ one has
$$
n_\a W^\a{}_{\b\m\n}e^\m_a \ell^\n=\nn y_{\b\m}e^\m_a -n_\b n_\a y^\a_\m e^\m_a +\frac{1}{m-1} {\cal G}_{\b\m}e^\m_a +\frac{H}{m} (\ell_a n_\b +\bar{g}_{ab}\omega^b_\b)
$$
from where
\bean
n_\a W^\a{}_{\b\m\n}e^\b_be^\m_a \ell^\n&=&\nn y_{ba}+\frac{1}{m-1} {\cal G}_{ba} +\frac{H}{m} \bar{g}_{ab}, \nonumber\\
n_\a  W^\a{}_{\b\m\n}\ell^\b e^\m_a \ell^\n &=&  \frac{1}{m-1} \ell^\b {\cal G}_{\b\m}e^\m_a -n_\a y^\a_\m e^\m_a +\frac{H}{m} \ell_a 
\eean
so that using (\ref{yG}) in the first or (\ref{ellGe}) in the second, as well as (\ref{trG})
\bea
n_\a W^\a{}_{\b\m\n}e^\b_be^\m_a \ell^\n &=&  \frac{2-m}{m-1} \left({\cal G}_{ab} - \frac{1}{m}{\cal G}^c{}_c \bar{g}_{ab}\right) \label{nWeeell} ,\\
n_\a  W^\a{}_{\b\m\n}\ell^\b e^\m_a \ell^\n &=& \frac{2-m}{m-1}\left(\ell_b {\cal G}^b{}_a - \frac{1}{m} {\cal G}^c{}_c\ell_a \right) \label{nWelleell}.
\eea
These provide a direct linear relation between these components of the Weyl singular part and the singular part of the  Einstein tensor.

The remaining independent components in the Weyl singular part are contained in 
\bea
\omega_\a^c W^\a{}_{\b\m\n}\ell^\b e^\m_a e^\n_b&=&\frac{1}{m-1} \left(\ell_a {\cal G}^c{}_b -\ell_b {\cal G}^c{}_a +\ell_d{\cal G}^d{}_a \delta^c_b - \ell_d {\cal G}^d{}_b\delta^c_a -\frac{2}{m}{\cal G}^d{}_d \left(\ell_a\delta^c_b -\ell_b \delta^c_a \right)\right)\label{omegaWee}\\
\omega_\a^d W^\a{}_{\b\m\n}e^\b_c e^\m_a e^\n_b&=& \frac{1}{m-1}\left(\bar{g}_{ac}{\cal G}^d{}_b - \bar{g}_{bc}{\cal G}^d{}_a-{\cal G}_{bc}\delta^d_a+{\cal G}_{ac}\delta^d_b -\frac{2}{m} {\cal G}^e{}_e \left(\bar{g}_{ac}\delta^d_b-\bar{g}_{bc}\delta^d_a \right)\right)\label{omegaWeee}
\eea
---these ones also determined by the Einstein singular part (and $\ell_a$)--- and also in
\be
\omega_\a^a W^\a{}_{\b\m\n}\ell^\b \ell^\m e^\n_b =y^a_b -\frac{1}{m-1}\left(\ell_b \omega^a_\a {\cal G}^\a_\m \ell^\m -{\cal G}^a{}_b \ll  -{\cal G}_{\b\m}\ell^\b\ell^\m \delta^a_b\right)+ \frac{H}{m} \delta^a_b \ll  \label{key}
\ee
which is a key equation, because I am going to show that it determines the {\em trace-free} part of $y_{ab}$ in terms of the Weyl and Einstein singular parts {\em independently of the causal character} of the shell $\S$. An alternative version of this relation is obtained multiplying by $\bar{g}_{ac}$ and using (\ref{omegaup}), (\ref{nWelleell}) and $\bar{g}_{ac}y^c_b =y_{ab} -\ell_a n^cy_{cb}$
\bea
e^\a_a W_{\a\b\m\n} \ell^\b \ell^\m e^\n_b &=& y_{ab} -\frac{1}{m-1} \left(\ell_b {\cal G}_{\a\m}\ell^\m e^\a_a +\ell_a {\cal G}_{\b\n}\ell^\b e^\n_b -\ll {\cal G}_{ab} -{\cal G}_{\b\m} \ell^\b \ell^\m \bar{g}_{ab}\right) \nonumber\\
&& +\frac{H}{m} \left(\ll \bar{g}_{ab} -\ell_a \ell_b \right) \nonumber \\
&=& y_{ab} \left(1-\frac{\ll\nn}{m-1} \right)-\frac{1}{m-1} \left(\ell_b n^cy_{ca}+\ell_a n^c y_{cb} +\bar{g}^{cd} y_{cd} \bar{g}_{ab} \right)
\nonumber\\
&& -\frac{2}{m(m-1)}(n^c n^d y_{cd}-\nn \bar{g}^{cd}y_{cd}) \left(\ll \bar{g}_{ab} -\ell_a \ell_b \right)\label{key1}
\eea
where in the second equality (\ref{G1}), (\ref{ellellG}), (\ref{trG}) and (\ref{ellGe}) have been used. By using
\be
\bar{g}^{ab} e^\a_a e^\b_b = g^{\a\b}-n^\a \ell^\b -n^\b \ell^\a +\nn \ell^\a \ell^\b \label{complete}
\ee
one easily checks that the lefthand side in (\ref{key1}) is identically trace-free (that is, its contraction with $\bar{g}^{ab}$ vanishes), and the righthand side too on using (\ref{yup}), (\ref{gngl}) and (\ref{gup}). Similarly, one can check that contracting (\ref{key1}) with $n^a$ and using (\ref{nWelleell}) one derives an identity. Hence, (\ref{key1}) contains no information concerning $\bar{g}^{cd}y_{cd}$ and no new information concerning $n^cy_{cd}$. Still, it carries information of the trace-free and $n^a$-orthogonal part of $y_{ab}$ and this information is essential for general shells, and in particular for everywhere null shells. 

In any case, the first equality in ({\ref{key1}) provides an explicit expression of $y_{ab}$ in terms of the singular parts of the curvature, which on using (\ref{ellGe}) and (\ref{ellellG}) reads
\bea
 y_{ab}= e^\a_a W_{\a\b\m\n} \ell^\b \ell^\m e^\n_b +\frac{1}{m-1} \left(2 \ell_c{\cal G}^c{}_{(a}\ell_{b)} -\ll {\cal G}_{ab} -\ell_c\ell_d {\cal G}^{cd}  \bar{g}_{ab}+\frac{2}{m}{\cal G}^d{}_d \left(\ll \bar{g}_{ab} -\ell_a \ell_b \right)\right) \label{main}
\eea
and holds for {\em general shells} everywhere.

This important relation adopts simplified forms in non-null, or null, points, if the corresponding preferred gauges are used. To see this, and 
as done with ${\cal G}_{\b\m}$, let us consider the implication of the above relations at null, and non-null, points of $\S$ separately.
\subsubsection{Non-null points}
At a {\em non-null} point $p\in\S$, and choosing there the canonical gauge, (\ref{nWelleell}) is empty, (\ref{nWeeell}) is equivalent to (\ref{key1}) and either of them is the trace of (\ref{omegaWeee}), while (\ref{omegaWee}) is exactly the same as (\ref{nWn}). To see what is the meaning of the remaining relations (\ref{nWn}) and (\ref{omegaWeee}), consider first the case of a shell which is spacelike at the considered point $p$, so that $\bm{n}|_p$ is chosen to be unit and timelike and we set $\epsilon |_p =-1$. Then, (\ref{nWn}) simply states that the {\em electric-magnetic} part of the Weyl singular distribution with respect to $\bm{n}|_p$ vanishes\footnote{For a brief recent exposition of the electric-magnetic decomposition of a Weyl-type tensor in arbitrary dimension see \cite{JSS}. In an orthonormal frame with `0' as the timelike index, the `electric-electric' part is given by the components $C_{0i0j}$, the `electric-magnetic' part by $C_{0ijk}$ and the `magnetic-magnetic' part by $C_{ijkl}$, see \cite{s-e,E-H}. The latter contains the electric-electric part due to the tracelessness of the Weyl tensor. The latter and the first parts are invariant under time reversal, while the second changes sign, see \cite{HOW}.}, while (\ref{omegaWeee}) expresses that the {\em magnetic-magnetic} part is pure trace, ergo fully equivalent to its {\em electric-electric} part, which is given by the traceless part of ${\cal G}_{ab}$ (expression (\ref{nWeeell})). This is actually always true in 4-dimensional spacetimes, but it is a non-trivial statement in higher dimensions. For points $q\in\S$ where the shell is timelike, an analogous identification follows, though now the terminology on `electric-magnetic' parts is not standard and might be misleading. Recall that, as seen before, at points where $\S$ is non-null the entire information carried by $y_{ab}$ is contained in ${\cal G}_{ab}$. This can be explicitly recovered from (\ref{main}) by using (\ref{nWeeell}) in the canonical gauge.

In summary, at any non-null point one has, in the canonical gauge
\be
W_{\a\b\m\n}|_p =\left. \epsilon 4n_{[\b}{\cal G}_{\a][\m}n_{\n]}-\frac{2}{m-1}\left({\cal G}_{\a[\m} g_{\n]\b}- {\cal G}_{\b[\m} g_{\n]\a}+2\epsilon {\cal G}^\r{}_\r n_{[\b}g_{\a][\m}n_{\n]} -\frac{2}{m}  {\cal G}^\r{}_\r  g_{\a[\m} g_{\n]\b} \right) \right|_p \, .
\ee

\subsubsection{Null points}
Consider now a point $s\in \S$ where $\S$ is {\em null}, so that $\nn |_s =0$. Then, relation (\ref{nWn}) states that $\vec n$ is a repeated principal null direction, or multiple WAND \cite{CMPP,MCPP,O}, of the tensor $W^\a{}_{\b\m\n}$, which is necessarily algebraically degenerate there, and of type II or more special ---in 4-dimensional $(M,g)$, this is Petrov type II. This was first noticed in \cite{BI}, see also \cite{BH}, for null shells in $m+1=4$. Observe that (\ref{nW}) simplifies now to
$$
n_\a W^\a{}_{\b\m\n}|_s =\left. \frac{2-m}{m-1}\left( n_\b (n_\a y^\a_{\n} n_{\m} -n_\a y^\a_{\m} n_{\n})+\frac{H}{2m} (g_{\b\m}n_{\n}-g_{\b\n}n_\m) \right)\right|_s
$$
where (\ref{G}) has been used, and now $H|_s =2 n^cn^d y_{cd}|_s$. Therefore, the algebraic type is more degenerate only if $H|_s =2 n^cn^d y_{cd}|_s=0$, so that $n_\a n_{[\g}W^\a{}_{\b]\m\n}|_s =0$, in which case the type is III (or more degenerate) \cite{O}. Observe that the scalar curvature has no singular part in this case, being just a (possibly discontinuous) function, and that ${\cal G}_{ab}|_s=0$ (but neither ${\cal G}^a{}_b|_s$ nor ${\cal G}^{ab}|_s$ vanish in this case, in general). If furthermore $n^\a y_{\a\b}|_s=0$, which is equivalent to $n^ay_{ab}|_s=0$, then the degeneracy is maximal with $n_\a W^\a{}_{\b\m\n}|_s =0$, the algebraic type is N \cite{O}, and $\vec n|_s$ is the unique WAND. In that situation, ${\cal G}^a{}_b|_s =0$ too, and 
\be
{\cal G}^{ab}|_s = - \bar{g}^{cd}y_{cd}\,  n^a n^b , \hs \mbox{(only for $W^\a{}_{\b\m\n}|_s$ of type N)}\label{GtypeN}
\ee
meaning that the Einstein tensor singular part takes the form of null radiation along $\vec n$ at these points $s\in \S$.

Relations (\ref{nWeeell}--\ref{omegaWeee}) just provide, at these null points, the lefthand side components of $W^\a{}_{\b\m\n}|_s$ in terms of surviving Einstein components, or equivalently, in terms of $n^ay_{ab}$. Observe, however, that at points $s$ where the algebraic type of $W^\a{}_{\b\m\n}|_s$ is III, then the components given in (\ref{nWeeell}) vanish, while at points where the algebraic type is N the components given in (\ref{nWelleell}--\ref{omegaWeee}) vanish too at $s$. There remains the important relation (\ref{key1}), which contains essential new information not included in the Einstein tensor. To extract this essential information, it is convenient to adopt a null gauge at $s$ choosing a null rigging $\ll|_s =0$ (so that $\ell_a \bar{g}^{ab}|_s=0$, $\ell_\m |_s=\ell_\a \omega^a_\m|_s$), and decompose $y_{ab}$ as
\be
y_{ab}|_s = \left.z_{ab} +n^cy_{ca} \ell_b +n^cy_{cb} \ell_a -\ell_a \ell_b n^c n^d y_{cd} +\frac{1}{m-1} \bar{g}_{ab}\,  \bar{g}^{cd} y_{cd} \right|_s
\label{ydecomp}
\ee
with
$$
n^c z_{cd} =0, \hs \bar{g}^{cd} z_{cd}=0.
$$
The symmetric traceless tensor $z_{ab}$ contains the sought information, and is transversal to the null generator $\vec n|_s$. It has $m(m+1)/2 -m-1 = (m+1)(m-2)/2$ independent components. Its spacetime version
$$
z_{\m\n} \defi z_{ab}\omega^a_\m \omega^b_\n
$$
is orthogonal to both $n^\m$ and $\ell^\m$ and traceless. Thus, it contains the typical information carried by a gravitational wave transversally to its null propagation direction. In 4-dimensional spacetimes these are just two degrees of freedom. This part of $y_{ab}$ never arises in the Einstein tensor singular part at null points, and thus it is purely gravitational, in the sense that is provided by the Weyl tensor distribution exclusively. The precise part of $W_{\a\b\m\n}$ that determines $z_{ab}$ is 
\be
{\cal W}_{ab}\defi (\delta^c_a -n^c \ell_a)(\delta^d_b-n^d\ell_b) e^\a_c W_{\a\b\m\n} \ell^\b \ell^\m e^\n_d  \label{w}
\ee
that is to say, the part of $W_{\a\b\m\n}$ with boost weight $-2$ in the null frame with $\{\vec n, \vec \ell\}$ as the two null vectors \cite{MCPP,CMPP}; in 4-dimensional spacetimes they are the two components encoded in the complex Newman-Penrose scalar $\Psi_4$ relative to $W_{\a\b\m\n}$. Hence, the essential new information contained in (\ref{key1}) can be written simply as
\be
{\cal W}_{ab}= z_{ab} \label{w=z}
\ee
at $s\in \S$, which can also be seen as a linear combination of (\ref{key1}) with (\ref{nWelleell}) and the trace of (\ref{nWeeell}) at $s$. 

Putting everything together, the expression of $y_{ab}$ in terms of the Einstein and Weyl singular parts is, at a null point $s\in \S$ and in a null gauge
\be
y_{ab}|_s =\left. {\cal W}_{ab}+\ell_c{\cal G}^c{}_a  \ell_b +\ell_c{\cal G}^c{}_b\ell_a -\frac{1}{m-1}\left(\ell_a \ell_b {\cal G}^c{}_c  +\bar{g}_{ab}\,  \ell_c\ell_d {\cal G}^{cd}\right) \right|_s \label{yW}.
\ee
This can also be easily recovered from the generally valid (\ref{main}) by using the null gauge, the definition (\ref{w}) and (\ref{nWelleell}). A relation equivalent to (\ref{yW}), but in arbitrary gauge, has been recently discussed in \cite{Marstalk}.

In summary, at null points $s\in \S$ one has that the $m(m+1)/2$ degrees of freedom given by the fundamental tensor $y_{ab}$ are codified as follows: its trace $\bar{g}^{ab}y_{ab}$ and the {\em longitudinal} (along $n^a$) components are given by the contravariant Einstein tensor singular part ${\cal G}^{ab}|_s$ according to (\ref{G3}) ---with $\nn|_s=0$. In General Relativity, these can be seen as energy-momentum degrees of freedom. The remaining $(m+1)(m-2)/2$ degrees of freedom $z_{ab}$ of $y_{ab}$ are encoded in the transversal and traceless Weyl components (\ref{w}), and they represent the typical transversal propagating degrees of freedom of a gravitational wave along the null direction $\vec n|_s$. The case where the longitudinal components are absent has a pure type N Weyl singular part, and a pure radiation Einstein tensor singular part (\ref{GtypeN}).

\section{Analysis of the shell field equations}\label{sec:analysis}
Now that the relation between the basic tensor $y_{ab}$ and the Einstein and Weyl singular parts of the curvature are known, the shell field equations can be given in terms of these singular parts, and thereby an analysis of their meaning and implications found.

I start by deducing the contracted parts of the field equations, because they are already known \cite{I,BI,MS,mars} and, more importantly, because they are usually the only ones taken into consideration. They can be obtained by taking traces of (\ref{eq1}-\ref{eq3}), but a quicker more efficient route is to take a double contraction of the mother relation (\ref{FE2}) itself. This leads to
\be
-n_\rho \left[G^\r_\l \right]=P_\rho^\sigma \nabla^\S_\sigma {\cal G}^\rho{}_\l +\varphi^\S_a \omega^a_\r {\cal G}^\r{}_\l \label{n[G]}
\ee
so that contracting here with $e^\l_b$ and using (\ref{P}), (\ref{nablabar}), (\ref{ellGe}) and (\ref{nG}) 
\be
-n_\r \left[G^\r_\l \right] e^\l_b = \Nb^\S_a {\cal G}^a{}_b +\varphi^\S_a {\cal G}^a{}_b +\ell_c{\cal G}^{ac}K^\S_{ab} . \label{div}
\ee
Similarly, contracting (\ref{n[G]}) with $\ell^\l$ and using (\ref{ellGomega}), (\ref{nablabar}) and (\ref{nablaell})
\be
-n_\rho \left[G^\r_\l \right]\ell^\l = \Nb^\S_a({\cal G}^{ab}\ell_b) -{\cal G}^a{}_b \Psi^b_{\S a}\label{div1}.
\ee
Equations (\ref{div}-\ref{div1}) provide a direct relation between the discontinuities of the Einstein tensor at $\S$ and the divergence of its singular part as a distribution. They reduce to the Israel equations \cite{I} for non-null shells in the canonical gauge, to the equations found in \cite{BI} for null shells, and they were more recently presented for general shells in \cite{mars}. They can actually be rewritten in a more convenient form by using the contravariant ${\cal G}^{ab}$. Raising the index $\l$ in (\ref{n[G]}) and then contracting with $n_\l$ and $\omega^b_\l$ one finds, respectively,
\bea
n_\r n_\l [G^{\r\l}]&=&{\cal G}^{ab}K^\S_{ab}, \label{nnG}\\
-n_\r [G^{\r\l}] \omega^b_\l &=& \Nb^\S_a {\cal G}^{ab} +\varphi^\S_a {\cal G}^{ab} \label{nGomega} .
\eea
In this form, they look much alike the original Israel equations for non-null shells, but they are valid for shells of arbitrary, even changing, causal character, and they can be checked to be fully equivalent to (\ref{div}-\ref{div1}). 
\begin{remark}
The ``extra'' term proportional to $\varphi^\S_a$ in (\ref{nGomega}) is actually the necessary term such that the righthand side of that equation provides a good divergence leading to conservation laws by integration. This follows because the rigged connection $\Nb^\S$ is not volume preserving in general \cite{Sch,MS}. Define a volume element $m$-form on $\S$ by
$$
\bar\eta_{a_1\dots a_m} \defi \ell^\r \eta_{\r \a_1\dots \a_m}e^{\a_1}_{a_1} \dots e^{\a_m}_{a_m}
$$
where $\eta_{\r \a_1\dots \a_m}$ is the canonical volume element $(m+1)$-form in $(M,g)$. This definition is independent of the choice of rigging ---but it depends on the choice of normalization for $\bm{n}$. A direct calculation shows that $\Nb^\S_b \bar\eta_{a_1\dots a_m}=\varphi^\S_b \bar\eta_{a_1\dots a_m}$, and this permits to check that, for any vector field $\vec v$ in $\S$, $d(i_{\vec v}\bar\eta )= (\Nb^\S_a v^a +\varphi^\S_a v^a)\bar\eta$, so that expression (\ref{nGomega}) is the correct one leading to conservation laws.
\end{remark}

Despite its simplicity and obvious interpretation, relations (\ref{nnG}) and (\ref{nGomega}) are just a small subset of the whole set of field equations on the shell, given by (\ref{eq1}--\ref{eq3}). For instance, (\ref{eq3}) can be rewritten, on using (\ref{yG}), like
\bea
-n_\a e^\b_c \left[R^\a{}_{\b\m\n}\right] e^\m_a e^\nu_b&=& \left(\Nb^\S_a +\varphi^\S_a \right)\left({\cal G}_{bc}-\frac{{\cal G}^d{}_d}{m-1} \bar{g}_{bc} \right)-\left(\Nb^\S_b +\varphi^\S_b \right)\left({\cal G}_{ac}-\frac{{\cal G}^d{}_d}{m-1}  \bar{g}_{ac} \right)\nonumber\\
&+&n^d\left(y_{ca} K^\S_{db} -y_{cb} K^\S_{da} - y_{db} K^\S_{ca}+y_{da} K^\S_{cb}\right)
\eea
which again is an expression valid for {\em general shells}. Here I have kept $y_{ab}$ in the second line, but an expression relating jumps of the curvature on the lefthand side with singular parts of the curvature on the righthand side is implicit by just replacing every instance of $y_{ab}$ by (\ref{main}). The resulting expression is long, but not more complicated. Similarly, substituting $y^a_b$ from (\ref{key}) into (\ref{eq2}), or alternatively $y_{ab}$ from (\ref{main}) into (\ref{eq2''}), one gets another set of field equations in terms of the curvature singular parts valid for arbitrary {\em general shells}, and analogously for (\ref{eq1}). 

Given the length of these equations in the general case, I am going to perform a detailed analysis for the cases of non-null, and null shells, bearing in mind that the results are applicable to just any general shell at particular points where $\S$ is non-null, or null, respectively, except maybe at points lying on a portion with empty interior of the closed set $\{s\in \S, \nn |_s=0\}$.

\subsection{The case of non-null shells}
Here, I adopt the canonical gauge. For the sake of readability I introduce the following Weyl projected tensors on $\S$ (on both sides)
\bea
E^\pm_{ab} &\defi& n^\a e^\b_a C^\pm_{\a\b\m\n} n^\m e^\n_b ,  \hs E^\pm_{ab}=E^\pm_{ba}, \hs \bar{g}^{ab}E_{ab}=0, \label{EE}\\
H^\pm_{abc} &\defi& n^\a e^\b_a C^\pm_{\a\b\m\n} e^\m_b e^\n_c , \hs H^\pm_{abc}=-H^\pm_{acb} ,\hs H^\pm_{[abc]} =0, \hs \bar{g}^{ab}H^\pm_{abc}=0 , \label{EH}\\
D^\pm_{abcd} &\defi & e^\a_a e^\b_bC^\pm_{\a\b\m\n} e^\m_c e^\n_d , \hs D^\pm_{abcd}=D^\pm_{[ab][cd]}, \hs D^\pm_{a[bcd]}=0, \hs \bar{g}^{ac}D^\pm_{abcd}= -\epsilon E^\pm_{bd}. \label{HH}
\eea
Observe that $D^\pm_{abcd}$ have the symmetries of a Riemann tensor, and thus themselves can be decomposed into traceless parts $F^\pm_{abcd}$ and their traces given by $E^\pm_{ab}$ such that
\bea
D^\pm_{abcd}=F^\pm_{abcd}-\frac{\epsilon}{m-2}\left(E^\pm_{bd} \bar{g}_{ac}-E^\pm_{bc} \bar{g}_{ad}-E^\pm_{ad} \bar{g}_{bc} + E^\pm_{ac} \bar{g}_{bd}\right), \label{F}\\
F^\pm_{abcd}=F^\pm_{[ab][cd]}, \hs F^\pm_{a[bcd]}=0, \hs \bar{g}^{ac}D^\pm_{abcd}=0 \hs \label{Fprop}
\eea
and here $F^\pm_{abcd}$ have the symmetries of a Weyl tensor, so that $F^\pm_{abcd}$ identically vanish if $m=3$. For the case of spacelike shells ($\epsilon =-1$) $E^\pm$, $H^\pm$ and $D^\pm$ represent the electric-electric, electric-magnetic, and magnetic-magnetic parts of the Weyl tensors $C^\pm$, see \cite{s-e,E-H,JSS}. I keep the same nomenclature for the timelike case too, though their interpretation is slightly different.

Using (\ref{nonnull1}-\ref{nonnull2}), the decomposition of the Riemann tensor in terms of the Weyl and Einstein tensors and (\ref{yG}), (\ref{eq2}) (or equivalently (\ref{eq2''})) becomes in this case 
$$
-[H^c{}_{ab}] -\frac{1}{m-1}\left([G_{\b\m}]n^\b e^\m_a \delta^c_b-[G_{\b\m}]n^\b e^\m_b \delta^c_a\right)=\Nb^\S_a {\cal G}^c{}_b-\Nb^\S_b {\cal G}^c{}_a+\frac{1}{2}\left(\delta^c_b \Nb_a H-\delta^c_a \Nb_b H \right)
$$
whose trace on $c$ and $a$ leads to
$$
\Nb^\S_c{\cal G}^c{}_b = [G_{\b\l}] n^\b e^\l_b
$$
which is nothing else but the Israel relation (\ref{nGomega}). Introducing this into the previous formula and using (\ref{trG}) one gets
\be
-[H^c{}_{ab}] =\Nb^\S_a {\cal G}^c{}_b-\Nb^\S_b {\cal G}^c{}_a+\frac{1}{m-1} \delta^c_b \left(\Nb^\S_d{\cal G}^d{}_a -\Nb_a{\cal G}^d{}_d \right)-\frac{1}{m-1} \delta^c_a \left(\Nb^\S_d{\cal G}^d{}_b -\Nb_b{\cal G}^d{}_d \right).\label{[H]}
\ee
This is one of the novel equations for non-null shells. It provides a relation between the jump of the Weyl part $H_{cab}$ and the derivatives of the Einstein tensor singular part. In general relativity, by using Einstein's field equations, it gives a relation between the derivatives of the energy-momentum tensor on the shell and the jump of $H_{cab}$. 

Consider now (\ref{eq1}) ---or equivalently (\ref{eq1''}). Using the same substitutions it becomes
\bea
\left[D^d{}_{cab} \right]+\frac{1}{m-1} \left([G^d{}_a]\bar{g}_{cb} - [G^d{}_b]\bar{g}_{ca} -[G_{ca}] \delta^d_b+[G_{cb}] \delta^d_a\right)+\frac{[R]}{m}\left(\delta^d_a \bar{g}_{cb}-\delta^d_b \bar{g}_{ca} \right)=\nonumber\\
\epsilon\left\{K^{d}_{\S a} \left({\cal G}_{cb}+\frac{H}{2} \bar{g}_{cb}\right)-K^{d}_{\S b} \left({\cal G}_{ca}+\frac{H}{2} \bar{g}_{ca}\right)+K^\S_{cb} \left({\cal G}^d{}_a +\frac{H}{2} \delta^d_a\right)-K^\S_{ca} \left({\cal G}^d{}_b +\frac{H}{2} \delta^d_b\right)\right\}\label{[D]}
\eea
and contracting here $d$ and $a$
\bea
&&-\epsilon [E_{cb}]-\frac{\epsilon}{m-1} n^\r n^\sigma[G_{\r\sigma}] \bar{g}_{cb} +\frac{m-2}{2m}[R]\bar{g}_{cb} +\frac{m-2}{m-1}[G_{bc}]=\nonumber\\
&&\epsilon\left\{K^d_{\S d} {\cal G}_{cb} -K^\S_{db}{\cal G}^d{}_c-K^\S_{dc}{\cal G}^d{}_b+K^\S_{cb} {\cal G}^d{}_d+\frac{H}{2}\left((m-2) K^\S_{cb}+\bar{g}_{cb}K^d_{\S d}\right)\right\} . \label{Edisc}
\eea
A further contraction here leads directly to 
$$
n^\r n^\sigma [G_{\r \sigma}] ={\cal G}^{cd} K^\S_{cd}
$$
which is (\ref{nnG}) for this case. Putting this into (\ref{Edisc}) 
\bea
&&-\epsilon [E_{cb}]+\frac{m-2}{m-1}\left([G_{bc}]-\frac{1}{m} \bar{g}_{bc}[G^d_d]\right)=\nonumber\\
&&\epsilon\left\{ K^d_{\S d} {\cal G}_{cb}-K^\S_{db}{\cal G}^d{}_c-K^\S_{dc}{\cal G}^d{}_b+\frac{1}{m-1} \left(K^\S_{cb} {\cal G}^d{}_d - K^d_{\S d} {\cal G}^e{}_e \bar{g}_{cb}  \right)+\frac{2}{m} K^\S_{ed}{\cal G}^{ed}\bar{g}_{bc} \right\} \label{[E]}
\eea
whose trace is identically vanishing. This is the second novel expression, relating the entire Einstein tensor singular part with the jump of the $E_{bc}$-part of the Weyl tensor and the traceless part of the jump of the tangential Einstein tensor {\em algebraically} ---together with intrinsic properties of the shell, i.e., the (mean) second fundamental form $K^\S_{ab}$.

The third and final novel relation is then found by introducing (\ref{[E]}) and (\ref{nnG}) into (\ref{[D]}) and after rearranging, leading to
\bea
\epsilon\left[ F_{dcab}\right] =2 K^\S_{d[a} {\cal G}_{b]c}-2 K^\S_{c[a} {\cal G}_{b]d}+\frac{4}{(m-1)(m-2)}\left(K^f_{\S f}{\cal G}^e{}_e -K^\S_{ef}{\cal G}^{ef}\right)\bar{g}_{d[a}\bar{g}_{b]c}\hs \hs \nonumber\\
-\frac{2}{m-2} \left\{K^e_{\S e}\left({\cal G}_{d[a}\bar{g}_{b]c}-{\cal G}_{c[a}\bar{g}_{b]d} \right) +{\cal G}^e{}_e \left(K^\S_{d[a}\bar{g}_{b]c}-K^\S_{c[a}\bar{g}_{b]d} \right)+2 K^\S_{e[c}\bar{g}_{d][b}{\cal G}^e{}_{a]}+2 K^\S_{e[b}\bar{g}_{a][c}{\cal G}^e{}_{d]}\right\} \label{[F]}
\eea
all of whose traces vanish identically. This is yet another algebraic relation between the Einstein tensor distribution singular part and the discontinuity of the $F$-part of the Weyl tensor. Observe that this relation is empty in 4-dimensional spacetimes ($m=3$), because then $F_{dcab}$ identically vanishes and the righthand side in (\ref{[F]}) also does due to a dimensionally dependent identity (see e.g. \cite{EH,EW}). However, in higher dimensions ($m>3$) this relation contains independent relevant information. 

Therefore, for non-null shells, the full set of field equations is given by the Israel standard equations together with equations (\ref{[H]}), (\ref{[E]}) and ---if $m>3$--- (\ref{[F]}). Observe that the latter two are purely algebraic, relating the algebraic structure of the second fundamental form with that of the Einstein tensor distribution singular part, while (\ref{[H]}) are differential conditions on ${\cal G}_{ab}$. Some of these relations have been used occasionally in the literature, specially concerning brane-worlds in 5-dimensional spacetimes ($m=4$), see \cite{SMS,MK}. However, the general restrictions that the new equations may pose on non-null shells and the additional properties that can be deduced from them remain to be analyzed in full generality.

\subsection{Null shells}
For everywhere null shells I adopt a null gauge with $\ll=0$ and then $\ell_a \bar{g}^{ab}=0$, $\ell_\m =\ell_\a \omega^a_\m$. It is convenient to introduce the following projector-like tensor field on $\S$
\be
\Pi^b_a \defi \delta^b_a -n^b \ell_a, \hs n^a\Pi^b_a =0, \hs \ell_b\Pi^b_a =0, \Pi^b_b =m-1, \hs \Pi^b_c\Pi^c_a =\Pi^b_a \label{Pi}
\ee
as well as the following vector fields
\be
E^\a_a \defi \Pi^b_a e^\a_b = e^\a_a -n^\a \ell_a, \hs n_\a E^\a_a=0, \hs \ell_\a E^\a_a =0.
\ee
Observe that
$$
n^a E^\a_a=0
$$
and therefore they constitute a set of $m-1$ linearly independent vector fields which, together with $n^\a =n^a e^\a_a$, form a basis of the tangent spaces in $\S$, and together with $\ell^\m$ a full basis of the tangent spaces $T_\S M$ at $\S$. The vector fields $\{\vec{E}_a\}$ thus represent the $m-1$ spacelike directions orthogonal to $n^\m$ and $\ell^\m$ on $\S$.

As in the previous case, I introduce the following Weyl projected tensors on $\S$
\bea
A^\pm_{ba}&\defi &n^\a E^\b_b n^\m E^\n_a C^\pm_{\a\b\m\n}, \hs A^\pm_{ba}=A^\pm_{ab}, \hs n^b A^\pm_{ba}=0, \hs \bar{g}^{ba} A^\pm_{ba} =0, \label{A}\\
B^\pm_{cab}&\defi &n^\a E^\b_c E^\m_a E^\n_b C^\pm_{\a\b\m\n}, \hs B^\pm_{cab}=-B^\pm_{cba}, \hs B^\pm_{[cab]}=0, \hs n^c B^\pm_{cab}=0, \, \, n^b B^\pm_{cab}=0, \label{B}\\
{\cal C}^\pm_{dcab} &\defi& E^\a_d E^\b_c E^\m_a E^\n_b C^\pm_{\a\b\m\n}, \hs {\cal C}^\pm_{dcab}={\cal C}^\pm_{[dc][ab]}, \hs {\cal C}^\pm_{d[cab]}=0, \, \,  n^d {\cal C}^\pm_{dcab}=0, \label{C}\\
L^\pm_{cab}&\defi& \ell^\a E^\b_c E^\m_a E^\n_b C^\pm_{\a\b\m\n}, \hs L^\pm_{cab}=-L^\pm_{cba}, \hs L^\pm_{[cab]}=0, \hs n^c L^\pm_{cab}=0, \, \, n^b L^\pm_{cab}=0, \label{L}\\
J^\pm_{cb} &\defi & \ell^\a n^\m E^\b_{c}  E^\n_{b} C^\pm_{\a\b\m\n}, \hs n^c J^\pm_{cb}=0, \hs n^b J^\pm_{cb}=0. \label{J}
\eea
Observe that, on using $\bar{g}^{ab} E^\a_a E^\b_b = \bar{g}^{ab} e^\a_a e^\b_b =g^{\a\b}-n^\a \ell^\b -n^\b  \ell^\a$, one has
\bea
\bar{g}^{cb}B^\pm_{cab}= -n^\a \ell^\b E^\m_a n^\n C^\pm_{\a\b\m\n}, \hs \bar{g}^{cb}L^\pm_{cab}= -\ell^\a n^\b E^\m_a \ell^\n C^\pm_{\a\b\m\n}, \label{trB}\\
\bar{g}^{da}{\cal C}^\pm_{dcab} =-2J^\pm_{(cb)}, \hs \bar{g}^{da}\bar{g}^{cb}{\cal C}^\pm_{dcab}= -2\bar{g}^{cb} J^\pm_{cb}= 2n^\a \ell^\b \ell^\m n^\n C^\pm_{\a\b\m\n}\label{trC}
\eea
still, the previous projected tensors do not exhaust all possible components of the Weyl tensors ---the components with boost weight $-2$ in the given null frame are missing, see \cite{CMPP}. Those given are just the needed ones in what follows. Note also that ${\cal C}^\pm_{dcab}$ can be seen as Riemann-like tensors in $m-1$ dimensions, and thus they can be further decomposed as (only for $m>3$)
\bea
{\cal C}^\pm_{abcd}=M^\pm_{abcd} -\frac{2}{m-3}\left(J^\pm_{(bd)} \bar{g}_{ac}-J^\pm_{(bc)} \bar{g}_{ad}-J^\pm_{(ad)} \bar{g}_{bc} + J^\pm_{(ac)} \bar{g}_{bd}\right)+\frac{4\bar{g}^{ef} J_{ef} }{(m-3)(m-2)} \bar{g}_{a[c}\bar{g}_{d]b}, \label{M}\\
M^\pm_{abcd}=M^\pm_{[ab][cd]}, \hs M^\pm_{a[bcd]}=0, \hs \bar{g}^{ac}M^\pm_{abcd}=0,\hs  n^a M^\pm_{abcd}=0. \hs \hs \label{Mprop}
\eea
As usual, $M^\pm_{abcd}$ identically vanish if $m=4$. In the important case with $m=3$ one simply has
$$
\mbox{if}\, \, \,  m=3 \hs \Longrightarrow \, \, \, {\cal C}^\pm_{abcd} = {\cal C} \left(\bar{g}_{ac}\bar{g}_{bd}-\bar{g}_{ad}\bar{g}_{bc}\right), \hs 2J_{(bd)} ={\cal C} \bar{g}_{bd}.
$$

The decomposition (\ref{yW}) holds now on the entire $\S$, that is
\be
y_{ab} = {\cal W}_{ab}+\ell_c{\cal G}^c{}_a  \ell_b +\ell_c{\cal G}^c{}_b\ell_a -\frac{1}{m-1}\left(\ell_a \ell_b {\cal G}^c{}_c  +\bar{g}_{ab}\,  \ell_c\ell_d {\cal G}^{cd}\right) \label{yW1}
\ee
with ${\cal W}_{ab}$ defined in (\ref{w}).

Consider the two main relations in their form given in (\ref{eq2''}) and (\ref{eq1''}). Using the decomposition of the Riemann tensor in terms of the Weyl and Einstein tensors, multiplication of (\ref{eq1''}) by $n^d$ leads to
\be
n^\a \left[C_{\a\b\m\n} \right]e^\b_ce^\m_a e^\n_b +\frac{2}{m-1}n^\a[G_{\a\m}]e^\m_{[a}\bar{g}_{b]c} =-2n^dy_{d[a}K_{b]c}\label{step}
\ee
and contracting here with $n^a$
$$
n^\a n^\m\left[C_{\a\b\m\n} \right]e^\b_c e^\n_b +\frac{1}{m-1} n^\a n^\m[G_{\a\m}] \bar{g}_{bc} =-K_{bc}\,  n^d n^ay_{da}
$$
whose trace with $\bar{g}^{cb}$ leads directly to (\ref{nnG}) via (\ref{trG}), and introducing this information into the equation one arrives at
\be
\left[A_{bc}\right]=\frac{1}{m-1} {\cal G}^d{}_d \left(K_{bc}-\frac{1}{m-1}\bar{g}^{ef}K_{ef} \bar{g}_{bc} \right) \label{[A]}
\ee
where the term in brackets on the righthand side is the {\em shear tensor} of the null hypersurface $\S$, that is, the trace-free part of the second fundamental form along $\vec n$ \cite{CSV}. The remaining information in (\ref{step}) is then (recall (\ref{ellGe}))
\be
\left[B_{cab}\right]+\frac{2}{m-1}n^\a[G_{\a\m}]E^\m_{[a}\bar{g}_{b]c} =-2\ell_d {\cal G}^d{}_{[a}K_{b]c} \label{[B]}
\ee
which can be rewritten without jumps of the Einstein tensor by using (\ref{div}). Equalities (\ref{[A]}) and (\ref{[B]}) contain $(m-2)(m-1)/2$ and $m(m-1)(m-2)/3$ independent relations, respectively. In 4-dimensional spacetimes the Weyl tensor jumps on the lefthand side of (\ref{[A]}) and (\ref{[B]}) correspond to the jumps of the complex Weyl scalars $\Psi_0$ and $\Psi_1$, respectively, and in this dimension these formulas were found in \cite{BH1}, see also \cite{BH}.

The rest of the information in (\ref{eq1''}) is then given by
\bea
 \left[{\cal C}_{dcab}\right] +\frac{2}{m-1} \left([G_{\a\m}]E^\a_d E^\m_{[a} \bar{g}_{b]c}+[G_{\a\m}]E^\a_c E^\m_{[b} \bar{g}_{a]d} \right) +\frac{2[R]}{m} \bar{g}_{d[a} \bar{g}_{b]c}=\nonumber\\
2 {\cal W}_{c[a}K_{b]d} + 2 K_{c[a}{\cal W}_{b]d}-\frac{2}{m-1} \ell^\b\ell^\m{\cal G}_{\b\m}\left(\bar{g}_{c[a}K_{b]d} +K_{c[a}\bar{g}_{b]d} \right). \label{[C]}
\eea
The cases with $m\in\{3,4\}$ are special as explained above. Thus, consider the trace of (\ref{[C]}), which reads in general
\bea
-2[J_{(cb)}] +\frac{1}{m-1} \left\{(m-3) [G_{\m\n}]E^\m_b E^\n_c -2[G_{\m\n}] n^\m\ell^\n \bar{g}_{bc} \right\}+\frac{m-4}{2m} [R]\bar{g}_{bc}=\nonumber\\\
-{\cal W}_{bc}\, \bar{g}^{de}K_{de}+K_{db} {\cal W}^d_{c}+K_{dc} {\cal W}^d_{b}+\frac{1}{m-1}\ell_a\ell_d{\cal G}^{ad}\left\{(m-3)K_{bc}+\bar{g}_{bc} \bar{g}^{de}K_{de} \right\} \label{[Jsym]}
\eea
and taking another trace here
\be
\bar{g}^{bc} [J_{bc}] +\frac{m-2}{m-1} 2[G_{\m\n}] n^\m\ell^\n+ \frac{m-2}{2m} [R]= -K_{de} {\cal W}^{de}-\frac{m-2}{m-1} \ell_a\ell_d{\cal G}^{ad}\, \bar{g}^{de}K_{de}  \label{[trJ]}
\ee
where we have used ${\cal W}^b_c = \bar{g}^{ba} {\cal W}_{ac}$ and ${\cal W}^{bc}=\bar{g}^{ca} {\cal W}^b_a$. Notice that only the the trace-free part of $K_{de}$, the shear tensor, enters in the first summand of the righthand side, while the second one is proportional to the trace of $K_{de}$, that is, to the {\em expansion} of the null generator along $\S$. For $m>3$ there are $m(m-1)/2$ independent relations in (\ref{[Jsym]}). If $m=3$ all the information in (\ref{[C]}), as well as in (\ref{[Jsym]}), is actually contained in (\ref{[trJ]}), which can be rewritten in this case simply as
$$
\mbox{if}\, \, \,  m=3 \hs \Longrightarrow \, \, \, [{\cal C}] +[G_{\m\n}] n^\m\ell^\n+\frac{1}{6} [R]=-K_{de}\left({\cal W}^{de}+\frac{1}{2} \ell_a\ell_d{\cal G}^{ad}\, \bar{g}^{de} \right).
$$
On the other hand, if $m=4$ all the information carried by (\ref{[C]}) is contained in (\ref{[Jsym]}). For the general case with $m>4$ (that is, 6-dimensional spacetimes or higher), one can use the decomposition (\ref{M}) and (\ref{[trJ]}) to get the information contained in (\ref{[C]}) but not included in (\ref{[trJ]}) as
\bea
\left[M_{dcab}\right]={\cal W}_{bd}K_{ca}-{\cal W}_{ad}K_{cb}-{\cal W}_{bc}K_{da}+{\cal W}_{ca}K_{bd}+\frac{2}{(m-2)(m-3)} {\cal W}^{ef}K_{ef}(\bar{g}_{da}\bar{g}_{bc}- \bar{g}_{db}\bar{g}_{ac})\nonumber\\
+\frac{2}{m-3}\left\{\bar{g}_{a[c}K_{d]e}{\cal W}^e_b +\bar{g}_{d[b}K_{a]e}{\cal W}^e_c+\bar{g}_{c[a}K_{b]e}{\cal W}^e_d+\bar{g}_{b[d}K_{c]e}{\cal W}^e_a-\bar{g}^{ef}K_{ef}\left(\bar{g}_{a[c}{\cal W}_{d]b}+\bar{g}_{b[d}{\cal W}_{c]a}\right)\right\}\label{[M]}
\eea
which is totally traceless. As already explained this collapses to a trivial identity $0=0$ if $m=4$ ---and is non-existent if $m=3$. For $m>4$ it contains $(m+1)m(m-1)(m-4)/12$ independent relations giving a direct algebraic connection between the discontinuity of the $M_{dcab}$ part of the Weyl tensor and the product of the second fundamental form with the ${\cal W}_{ab}$-term in the singular part of the Weyl tensor distribution. 

Resort now to the relation (\ref{eq2''}). First, I note that multiplying there by $n^a n^c$ the following results
$$
n^\a\ell^\b\left[C_{\a\b\m\n} \right]n^\m E^\n_b -\frac{1}{m-1}n^\a[G_{\a\n}]E^\n_b=n^c\Nb^\S_c\left(\ell_d{\cal G}^d{}_b+\frac{H}{2}\ell_b \right)+2\ell_d{\cal G}^{cd}K_{cb}-\frac{1}{2}\vec{e}_b(H) -\varphi^\S_b \frac{H}{2}
$$
which combined with (\ref{trB}) and the trace of (\ref{[B]}) leads directly to the Israel relation (\ref{div}) projected with $\Pi^b_c$, and therefore can be omitted. Contracting (\ref{eq2''}) just with $n^a$ and projecting leads in turn to
\bea
[J_{cb}]+\frac{1}{m-1}\left([G_{\m\n}]E^\m_cE^\n_b +\bar{g}_{cb}[G_{\m\n}]\ell^\m n^\nÊ\right)+\frac{1}{m}[R]\bar{g}_{bc} =
\Pi^d_b \Pi^e_c \left(\Nb^\S_d(n^ay_{ae}) -n^a\Nb^\S_ay_{de} \right)\nonumber\\
+n^a \varphi^\S_a \left({\cal W}_{bc}-\frac{1}{m-1} \bar{g}_{bc} \ell_d\ell_e{\cal G}^{de} \right)-K_{eb}{\cal W}^e_c +\frac{1}{m-1}K_{bc} \ell_d\ell_e{\cal G}^{de} \label{[J]}
\eea
which is convenient to split into its symmetric and skew-symmetric parts. Again, note that the Weyl jump on the lefthand side in (\ref{[J]}) corresponds to the jump of the Weyl scalar $\Psi_2$ if $m=3$, and the symmetric and anty-symmetric parts of the former to the real and imaginary parts of the latter. In this special instance of 4-dimensional spacetimes expression (\ref{[J]}) was written down in \cite{BH1} for the special case of ``null shells with a type III geometry'', \cite{P}, which essentially amounts to assuming $n^cy_{bc}=0$, or equivalently $\ell_b{\cal G}^b{}_c=0=H$. The rationale behind this restriction is unclear to me, and there seems to be no reason in principle why such a restriction should be enforced. 

Concerning the anty-symmetric part of (\ref{[J]}) one obtains
$$
[J_{[cb]}]=\Pi^d_{[b} \Pi^e_{c]} \Nb^\S_d(n^ay_{ae})+K_{e[b}{\cal W}^e_{c]} 
$$
that can be easily written as
\be
[J_{[cb]}]=\Pi^d_{[b} \Pi^e_{c]} \Nb^\S_d(\ell_f{\cal G}^f{}_e) +{\cal G}^f{}_f \Pi^d_{[b} \bar{g}_{c]e}\Psi^e_{\S d}+K_{e[b}{\cal W}^e_{c]} \label{[Jasym]}.
\ee
There are $(m-1)(m-2)/2$ independent relations here.
On the other hand, the symmetric part of (\ref{[J]}) can be combined with (\ref{[Jsym]}) to arrive, after a little calculation, at
\bea
[G_{\m\n}]E^\m_cE^\n_b +\frac{1}{2}[R]\bar{g}_{bc}=2\Pi^d_{(b} \Pi^e_{c)} \left(\Nb^\S_d(n^ay_{ae}) -n^a\Nb^\S_ay_{de} \right)+\nonumber\\
(2n^a \varphi^\S_a -\bar{g}^{af}K_{af})\left({\cal W}_{bc}-\frac{1}{m-1}\ell_d\ell_e{\cal G}^{de}\,  \bar{g}_{bc}  \right)+K_{bc}\ell_d\ell_e{\cal G}^{de}.\label{[Ricci]1}
\eea
Observe that the lefthand side can be written in terms of the Ricci tensor exclusively. The trace of this relation gives (\ref{div1}), so that only the traceless part provides new information. 
After another calculation this formula can be expressed as
\bea
[R_{\m\n}]E^\m_cE^\n_b= [G_{\m\n}]E^\m_cE^\n_b +\frac{1}{2}[R]\bar{g}_{bc}=-2n^a\Nb^\S_a {\cal W}_{bc} +{\cal W}_{bc} (2n^a \varphi^\S_a -\bar{g}^{af}K_{af})\nonumber\\
- \ell_d\ell_e{\cal G}^{de} \left(K_{bc} -\frac{1}{m-1} \bar{g}^{af}K_{af} \bar{g}_{bc} \right)
-2n^e\Psi^f_{\S e}\ell_a \left({\cal G}^a{}_{(b}\bar{g}_{c)f}-\frac{1}{m-1}{\cal G}^a{}_{f}\bar{g}_{bc}\right)
\nonumber\\
+ 2\bar{g}_{a(c}\Pi^d_{b)} \Nb^\S_d(\ell_f{\cal G}^{fa}) -2{\cal G}^e{}_{(b}\bar{g}_{c)f}\Psi^f_{\S e}
+\frac{2}{m-1} \ell_e n^a\Nb^\S_a (\ell_d{\cal G}^{de}) \bar{g}_{bc} \label{[Ricci]}.
\eea
This is an important formula providing the information contained in (\ref{div1}) plus $(m+1)(m-2)/2$ independent new relations in its traceless part. This traceless part is the evolution equation (along the null shell) for the portion of the Weyl tensor singular part ${\cal W}_{ab}$, which represents a propagating transversal gravitational wave, in terms of Ricci (or Einstein) tensor discontinuities. In 4-dimensional spacetimes a restricted version (to `type III' shells) appear in \cite{BH,BH1}. Evolution equations equivalent to (\ref{[Ricci]}), in gauge covariant form and in terms of the hypersurface metric connection \cite{mars}, appears in \cite{Marstalk} for the general dimensional case.

At this stage the only information remaining is the projected part of (\ref{eq2''}), which after a similar calculation reads
\bea
\left[ L_{cab}\right]+\frac{2}{m-1} \ell^\m[G_{\m\n}]E^\n_{[a} \bar{g}_{b]c}=
2\Pi^e_{[b}\Pi^d_{a]}\Pi^f_c \Nb^\S_e {\cal W}_{df}+2\varphi^\S_d \Pi^d_{[a} {\cal W}_{b]c}-\frac{2}{m-1} \ell_d \ell_e {\cal G}^{de}\varphi^\S_{[a}\bar{g}_{b]c} \nonumber \\
+2\Psi^f_{\S e} \Pi^e_{[b} \left(\bar{g}_{a]f}{\cal G}^d{}_c+{\cal G}^d{}_{a]}\bar{g}_{cf}\right)\ell_d 
-\frac{2}{m-1} \bar{g}_{c[a} \Pi^ e_{b]} \Nb^\S_e\left(\ell_d\ell_f{\cal G}^{df} \right).\label{[L]}
\eea
On the lefthand side, the Einstein tensor jumps $\ell^\m[G_{\m\n}]E^\n_b$ appear together with those of the  the Weyl tensor components of boost weight $-1$. In the case $m=3$ the latter correspond to jumps in the $\Psi_3$ Weyl scalar. In general, there are $m(m-1)(m-2)/3$ independent relations in (\ref{[L]}). As far as I am aware, this relation was previously unknown, even in 4-dimensional spacetimes.

\subsubsection{Summary of field equations for null shells}
Summarizing this subsection, the full set of field equations for null shells is given by the Israel formulas (\ref{nnG}-\ref{nGomega}) together with (\ref{[A]}), (\ref{[B]}), (\ref{[Jsym]}), (\ref{[M]}), (\ref{[Jasym]}), the trace-free part of (\ref{[Ricci]}), and (\ref{[L]}). The following remarks are in order:
\begin{itemize}
\item As was already known, for null shells (\ref{nnG}-\ref{nGomega}) do not involve the full set of $m(m+1)/2$ components of $y_{ab}$. These are encoded in $\ell_b{\cal G}^{bc}$ ($m$ components; alternatively, in $\ell_b{\cal G}^b{}_c$ plus $\ell_b\ell_c {\cal G}^{bc}$) and  ${\cal G}^b{}_b$ ($1$ component) together with the traceless transversal Weyl singular part ${\cal W}_{ab}$ ($(m+1)(m-2)/2$ components). The decomposition (\ref{yW1}) gives the precise relation between them in a null gauge. The Israel equations (\ref{nnG}-\ref{nGomega}) rule only $\ell_b{\cal G}^{bc}$ and ${\cal G}^b{}_b$, but say nothing about ${\cal W}_{ab}$.
\item The jumps of the boost-weight $-2$ components do not arise in the equations, neither for the Weyl tensor nor for the one in the Einstein tensor, $\ell^\m \ell^\n [G_{\m\n}]$. Thus, they have no influence on the shell.
\item Eqs.  (\ref{[A]}), (\ref{[B]}), (\ref{[Jsym]}) and (\ref{[M]}), as written, are purely algebraic in the sense that they do not involve any derivatives of the variables $\ell_b{\cal G}^{bc}$, ${\cal G}^b{}_b$ or ${\cal W}_{ab}$. Therefore, in a way they have a similar status to the Israel condition (\ref{nnG}).
\item In those `algebraic' equations, an essential ingredient entering on every term of their righthand sides is the second fundamental form of the shell $K_{ab}$, which is known to be intrinsic to null hypersurfaces and related to the Lie derivative of the (degenerate) first fundamental form along the null generator. It follows that for totally geodesic shells (that is, with $K_{ab}=0$, see \cite{Mars2012} and references therein), the jumps on the lefthand sides must all vanish. This applies for instance to shells on non-expanding horizons and on Killing horizons (e.g., the general Killing-horizon shells inside vacuum solutions of Einstein equations built in \cite{BO}), and also to impulsive plane waves traveling on a flat spacetime. 
\item The remaining relations (\ref{[Jasym]}), (\ref{[Ricci]}) and (\ref{[L]}) are differential equations for the variables  $\ell_b{\cal G}^{bc}$ and ${\cal W}_{ab}$, and thus comparable to the Israel equation (\ref{nGomega}). Eq.(\ref{[Ricci]}) gives the {\em evolution}, along the null generators $\vec n$, of the mentioned variables. The evolution properties of $\ell_b{\cal G}^{bc}$ are described by (\ref{nGomega}) ---which is the trace of (\ref{[Ricci]})---, while the evolution of ${\cal W}_{ab}$ is given by the traceless part of (\ref{[Ricci]}). The latter is thus essential to consider the evolution system complete.
\item Eqs.(\ref{[Jasym]}) and (\ref{[L]}) are also differential, but of a different kind, as they involve derivatives tangent to $\S$ but {\em transverse} to the null generators. In this sense, they can be considered to be {\em constraints} on any possible initial data (for their evolution equations (\ref{[Ricci]})) one may wish to impose on the variables $\ell_b{\cal G}^{bc}$ or ${\cal W}_{ab}$. Equation (\ref{[Jasym]}) provides these ``tangent to $\S$ but transversal to $\vec n$'' derivatives for the Einstein variables $\ell_b{\cal G}^b{}_c$ while (\ref{[L]}) provides those of the Weyl part ${\cal W}_{ab}$ and of $\ell_b\ell_c{\cal G}^{bc}$.
\end{itemize}

The importance of the novel part of the full set of field equations for null shells, as well as the general restrictions that they may impose, remain to be studied in depth.

\section{Concluding remarks}\label{Conclusions}
In this paper the full set of field equations for thin shells of arbitrary, even changing, causal character have been derived in arbitrary spacetime dimensions. The two prominent cases of everywhere non-null, and everywhere null, shells have been analyzed in higher depth. New equations relating the jumps of the curvature on the ambient manifold with the singular parts of the Einstein and Weyl tensor  distributions ---and the intrinsic properties of the shell--- have been presented. 

To end, I would like to stress two significant features:
\begin{enumerate}
\item A first important remark is that the contents of this paper can be used as a {\em theoretical framework} to study thin shells in general gravitational theories. The reason is that all the formulas presented have a purely geometrical character, and therefore are independent of, and valid in, any particular gravitational theory based on a Lorentzian manifold. To apply them to any selected such theory, one only needs to add the corresponding field equations and analyze the restrictions that they may impose on the curvature tensor distribution and its distributional derivatives, if any. 

As an example, take the prominent case of General Relativity, where the field equations read (for tensor distributions)
\be
\underline{G}_{\m\n} +\Lambda \underline{g}_{\m\n}=\kappa \underline{T}_{\m\n} \label{efe}
\ee
where $\underline{T}_{\m\n}$ is the energy-momentum tensor distribution, $\Lambda$ a cosmological constant\footnote{The cosmological constant takes, in general, the form $\Lambda =\Lambda^+ \theta +\Lambda^- (1 -\theta)$ so that it may take different values at both sides of the shell $\S$, see \cite{MSV}.} and $\kappa =c^4/(8\pi G)$ is the gravitational coupling constant. The structure of the lefthand side induces automatically a structure for $\underline{T}_{\m\n}$ like
$$
\underline{T}_{\m\n} = T^+_{\m\n} \otheta +T^-_{\m\n} (\1-\otheta) +\tau_{\m\n} \deltaS
$$
where $\tau_{\m\n}\deltaS$ is the singular part with support on $\S$ of $\underline{T}_{\m\n}$. Then, the Einstein field equations (\ref{efe}) imply directly
$$\tau_{\m\n} ={\cal G}_{\m\n}/\kappa $$
and all the formulas involving ${\cal G}_{\m\n}$ are automatically valid for $\kappa \tau_{\m\n}$. 

A more elaborated example is given by the most general quadratic theory, whose field equations can be found, e.g., in \cite{RSV}. Then, the existence of products of the Riemann and Ricci tensor distributions in these field equations implies that the curvature cannot carry a singular part ---for the product of distributions is ill-defined. Therefore, one must impose that the entire $H^\a{}_{\b\m\n}$ in (\ref{Hriemann}) vanishes ---except for the particular case where the theory has a Lagrangian $R+\alpha R^2$, \cite{RSV,Senovilla13}. This readily implies that $y_{ab}=0$, and thereby that ${\cal G}_{\m\n}$ and $W^\a{}_{\b\m\n}$ vanish too. In order to get the expression of the energy-momentum singular part then a calculation involving the distributional derivatives of $R^\a{}_{\b\m\n}$ is needed. See \cite{RSV,Senovilla13} for the case of non-null shells, where one comes upon the existence of {\em gravitational double layers}, that is to say, shells in which the singular part of the energy-momentum tensor $\underline{\tau}_{\m\n}$ involves not only a term proportional to  $\deltaS$, but also a term proportional to its `derivative' \cite{S2,S3,RSV}.

\item In the same vein, the formulas in this paper can be used to derive the {\em proper matching conditions} on general hypersurfaces, that is to say, the conditions such that $\S$ does not support any distributional energy-momentum, which will at most have discontinuities across $\S$. This is the case of the world-surface of a star or finite object in general, or of shock matter waves. To that end, one simply considers the selected field equations and computes the conditions such that $\underline{T}_{\m\n}$ is actually a well-defined ---though maybe discontinuous across $\S$--- tensor field.

Take as fundamental example General Relativity again. Then from the discussion above we know that $\tau_{\m\n}=0$ if and only if ${\cal G}_{\m\n}=0$, so that the latter has to be imposed in all the equations and in particular in (\ref{main}). The derived conditions on discontinuities of the curvature were found in \cite{MS} for general shells. For non-null shells this leads to the vanishing of $W^\a{}_{\b\m\n}$ too, and the equations only provide the allowed discontinuities of the curvature. For null shells, though, the ${\cal W}_{ab}$-part of $W^\a{}_{\b\m\n}$ survives in principle, allowing for the possibility that an impulsive gravitational wave travels along the shell. This propagation is then ruled by the field equation (\ref{[Ricci]}), and ${\cal W}_{ab}$ itself is constrained by the relations (\ref{[Jsym]}), (\ref{[M]}), (\ref{[Jasym]}) and (\ref{[L]}) appropriately restricted with ${\cal G}_{ab}=0$.

The conditions imposed by the vanishing of the singular part of $\underline{T}_{\m\n}$ in other gravitational theories may be more involved. The cases of non-null shells in $F(R)$-gravity, and in general quadratic gravity, can be consulted in \cite{Senovilla13} and \cite{RSV}, respectively.

\end{enumerate}

As a final comment, I would like to stress that if the matter contents of the problem under consideration is identified, then the corresponding field equations and junction conditions for the fields involved must be added to the picture.

\section*{Acknowledgments}
I thank Marc Mars for reading the manuscript and for informing of the existence of the results in \cite{Marstalk}, as well as for providing a copy of the draft. I am also grateful to Kepa Sousa and Ra\"ul Vera for interesting comments and suggestions that improved previous versions of the manuscript. 
Support from Grants No. FIS2017-85076-P (Spanish MINECO-fondos FEDER) and No. IT956-16 (Basque Government) is acknowledged.

\appendix

\section{Appendix}

Given $\S$, $\{\vec\ell,\vec{e}_a\}$, $\{\bm{n},\bm{\omega}^a\}$, and $\bar{g}$ as explained in section \ref{sec:basic}, then the following formulas, objects and notations are used \cite{MS}
\bea
\ell_a := \ell_\mu e^\mu_a, \hs \ell_\mu = \ell_a \omega^a_\mu +\ll n_\mu, \label{ella}\\
n^a \defi n^\mu \omega^a_\mu \hs n^\mu = n^a e^\mu_a +\nn \ell^\mu, \label{na}\\
n^a\ell_a = 1-\nn \ll , \label{nala}\\
\bar{g}^{ab} \defi g^{\mu\nu} \omega_\mu^a \omega_\nu^b , \hs \bar{g}^{ab}\bar{g}_{bc} =\delta^a_c -n^a\ell_c ,\label{gup}\\
\bar{g}_{ab} n^b = -\nn \ell_a , \hs \bar{g}^{ab}\ell_b = -\ll n^a ,\label{gngl}\\
g^{\mu\nu} \omega_\nu^a= n^a \ell^\mu+\bar{g}^{ab} e^\mu_b, \hs g_{\mu\nu}e^\nu_a =\ell_a n_\mu +\bar{g}_{ab} \omega^b_\mu , \label{omegaup}\\
P^\mu_\nu \defi \delta^\mu_\nu -\ell^\mu n_\nu = e^\mu_a \omega^a_\nu, \hs P^\mu_\nu \ell^\nu =0, \hs P^\mu_\nu n_\mu =0\label{P}
\eea
The Levi-Civita connections on $M^\pm$ induce corresponding torsion-free {\em rigged} connections $\bar{\Gamma}^\pm$ on $\S$ defined by
\be
\omega^c_\rho e^\mu_a\nabla^\pm_{\mu} e^\rho_b  \defi \bar{\Gamma}^{\pm c}_{ab} \label{riggedGamma}
\ee
and the respective covariant derivatives are denoted by $\Nb^\pm$. For the derivatives of the normal and the rigging one defines \cite{MS}
\bea
K^\pm_{ab} &\defi& e^\mu_a e^\nu_b\nabla^\pm_\mu n_\nu =K^\pm_{(ab)}, \label{K}\\
\Psi^a_{\pm b} &\defi& \omega^a_\mu e^\nu_b\nabla^\pm_\nu \ell^\mu, \label{Psi}\\
\varphi^\pm_a &\defi& n_\mu e^\nu_a \nabla^\pm_\nu \ell^\mu, \label{phi}\\
\H^\pm_{ab} &\defi& e^\mu_a e^\nu_b\nabla^\pm_\mu \ell_\nu. \label{H}
\eea
For an arbitrary tensor field $T^\mu_\nu$ with well-defined limit on $\S$ one then has on $\S$
\bea
\omega^a_\mu e^\rho_b e^\nu_c\nabla^\pm_\nu T^\mu_\rho = \Nb^\pm_c \bar{T}^a_b +\Psi^a_{\pm c} n_\mu T^\mu_\rho e^\rho_b+ K^\pm_{bc} \omega^a_\mu T^\mu_\rho \ell^\rho,\\
\omega^a_\mu e^\rho_b e^\nu_c\nabla^\S_\nu T^\mu_\rho = \Nb^\S_c \bar{T}^a_b +\Psi^a_{\S c} n_\mu T^\mu_\rho e^\rho_b+ K^\S_{bc} \omega^a_\mu T^\mu_\rho \ell^\rho \label{nablabar}
\eea
and its natural extension to an arbitrary number of indices. Here $\bar{T}^a_b := \omega^a_\mu T^\mu_\nu e^\nu_b$ is the projection of the tensor $T$ to $\S$ according to the chosen bases, and $\nabla^\S, \overline\nabla^\S$ are defined by
$$
\nabla^\S_\mu \defi \frac{1}{2} \left(\nabla^+_\mu + \nabla^-_\mu\right), \hs 
\overline\nabla^\S_c \defi \frac{1}{2} \left(\overline\nabla^+_c + \overline\nabla^-_c\right).
$$
Hence
\bea
e^\m_a\nabla^\pm_\m n_\n = -\varphi^\pm_a n_\n +K^\pm_{ab}\omega^b_\n, \hs \
\vec{e}_a\nn =2K^\S_{ab}n^b -2\varphi^\S_a \nn \label{nablann}\\
e^\m_a\nabla^\pm_\m \ell^\n =\varphi^\pm_a \ell^\n +\Psi^b_{\pm a}e^\n_b, \hs
\vec{e}_a \ll = 2\ell_b\Psi^b_{\S a}+2\varphi^\S_a \ll, \label{nablall}
\eea
In particular one also has
\bea
\Nb^\pm_c \bar{g}_{ab} &=&-\ell_b K^\pm_{ac}-\ell_a K^\pm_{bc}, \label{nablag}\\
\Nb^\pm_c \bar{g}^{ab}&=& -n^b \Psi^a_{\pm c} - n^a \Psi^b_{\pm c}, \label{nablagup}\\
\Nb^\pm_a n^b &=& \bar{g}^{bc}K^\pm_{ca} -\varphi^\pm_a n^b -\nn \Psi^b_{\pm a}, \label{nablan}\\
\Nb^\pm_b \ell_a &=& -\ll K^\pm_{ab} +\bar{g}_{ac} \Psi^c_{\pm b} +\ell_a \varphi^\pm_b .\label{nablaell}
\eea
and their corresponding versions substituing the $\pm$ by $\S$.
Observe that $\H_{ab}$ and $\Psi^b_a$ are not independent:
\be
\H^\pm_{ba} = \bar{g}_{ac} \Psi^c_{\pm b} +\ell_a \varphi^\pm_b \label{HPsi}
\ee
so that one can rewrite (\ref{nablaell}) as
$$
\Nb^\pm_b \ell_a = -\ll K^\pm_{ab} +\H^\pm_{ba}
$$ 
proving in particular that the skew-symmetric parts $\H^+_{[ab]}=\H^-_{[ab]}$ always coincide on $\S$. It is therefore useful to introduce the following symmetric objects \cite{mars}
\be
Y^\pm_{ab}:= \H^\pm_{(ab)} = \Nb^\pm_{(a} \ell_{b)} + \ll K^\pm_{ab}.\label{Y}
\ee
$Y_{ab}$ is a fundamental, {\em rigging-dependent},  object that measures the departure of the shell $\S$ from a standard hypersurface, in the sense that $\S$ is not a shell if and only if the tensors $\H^\pm_{ab}$ agree from both sides $M^\pm$. It is thus convenient to define its jump on $\S$
\be
y_{ab} \defi \left[ \H_{ab} \right] = \left[ \H_{(ab)} \right] = \left[ Y_{ab} \right] \label{y}
\ee
and its spacetime version
\be
y_{\mu\nu} \defi y_{ab} \omega^a_\mu \omega^b_\nu , \hs \ell^\mu y_{\mu\nu} =0.\label{y1}
\ee
From (\ref{Y}) it is immediate that
\be
y_{ab}= \left[ \Nb_{(a}\ell_{b)}\right] +\ll \left[ K_{ab}\right] \label{ygen}
\ee
which is valid for general shells.
A fundamental result is that $y_{ab}$ does not depend on the rigging choice \cite{MS}.
Then one can prove \cite{MS}
\be
\left[\Psi^a_b \right]=\bar{g}^{ac}y_{cb}, \hs \left[ \varphi_a\right]=n^b y_{ab}, \hs \left[K_{ab}\right] =\nn y_{ab},\hs \left[ \bar{\Gamma}^a_{bc}\right]=-n^a y_{bc} \label{jumps}
\ee
as well as
\be
\left[ \Gamma^\a_{\mu\nu} \right]= y^\a_\mu n_\nu +y^\a_\nu n_\mu - n^\a y_{\mu\nu}.\label{Gammajump}
\ee

In our situation, the spacetime $(M,g)$ contains a general {\em shell}, that is, a hypersurface $\Sigma$ such that the metric tensor $g$ may be not differentiable, if $y_{ab}\neq 0$, across $\S$. Nevertheless, given that the metric tensor $g$ is continuous, it defines a tensor distribution $\underline g$ \footnote{Following \cite{MS,RSV}, I will usually put an underline on distributional objects to emphasize their distributional character.} which can be differentiated {\em in the distributional sense}. For the basics on tensor distributions and their covariant derivatives the reader is referred to \cite{MS,RSV,GT,L,L1,T,SV}. In particular, the curvature can be defined as a tensor distribution. First of all, define the $\Sigma$-step function $\theta : M \rightarrow  \mathbb{R} $ as
\bea
\theta =\left\{
\begin{array}{ccc}
1 &  & M^+\\
1/2 & \mbox{on} & \Sigma \\
0 &  & M^-
\end{array}\right. \label{theta}
\eea
and observe that $\theta^\S =\theta|_\S=1/2$ in agreement with our definition of the value of discontinuous objects at $\S$. $\theta$ defines a scalar distribution $\otheta$ whose covariant derivative is a one-form distribution with support on $\S$ denoted by
\be
\underline{\bm{\d}} := \nabla \otheta \label{deltamu}
\ee
which can be easily seen to be collinear with $\bm{n}$ \cite{MS}. Thus, one can write
\be
\underline{\bm{\d}} =\bm{n} \deltaS \label{delta}.
\ee
It thus arises a scalar distribution $\deltaS$ with support on $\S$. It must be remarked, however, that for general shells containing null points, $\deltaS$ is not univocally defined, as it depends on the normalization factor of the normal one-form $\bm{n}$ which itself is not canonically fixed. Fixing it would require choosing a definite volume element on $\S$ (this can be easily done for everywhere timelike, or spacelike, hypersurfaces, but not for $\S$ with null points). Therefore, one should keep in mind that only the product (\ref{delta}) of $\deltaS$ with $n_\mu$ is well defined, even though the use of $\deltaS$ is very useful in many occasions.

Let $T$ denote any $(p,q)$-tensor field which may be discontinuous across $\Sigma$ but with 
definite limits on $\S$ from $M^\pm$. In accordance with (\ref{TSigma}) $T$ can be expressed, as a tensor field and as a distribution, as
\be
T= T^{+} \theta +T^{-} \left( 1-\theta \right ), \hs \underline{T}=T^{+} \otheta +T^{-} \left( \1-\otheta \right ) \enspace . \label{Tdist}
\ee
The covariant derivative of $\underline T$ is \cite{MS,RSV}
\be
\underline{\nabla T} = \nabla T^+ \otheta +\nabla T^- (\1 -\otheta) +\underline{\bm{\delta}}\otimes \left[T\right]  \label{nablaT}
\ee
with $\left[ T\right ]$ as defined in (\ref{discont}), or with indices
\be
\nabla_\mu\, \underline{T}^{\a_1\dots\a_q}_{\b_1\dots\b_p} = \nabla_\mu T^{+\a_1\dots\a_q}_{\b_1\dots\b_p} \otheta
+\nabla_\mu T^{-\a_1\dots\a_q}_{\b_1\dots\b_p} (\1 -\otheta )+\left[T^{\a_1\dots\a_q}_{\b_1\dots\b_p} \right] n_\mu \deltaS.
\label{nablaT1}
\ee

Using for the continuous metric
\begin{eqnarray*}
g= g^{+}  \theta+g^{-} \left( 1- \theta \right) , \hspace{1cm} \underline{g}=g^{+}  \otheta + g^{-} \left( \1- \otheta \right) 
\end{eqnarray*}
a standard calculation provides the curvature as a tensor distribution, given by \cite{MS,RSV}
\be
\underline{R}^\alpha{}_{\beta\mu\nu}=R^{+\alpha}{}_{\beta\mu\nu}\otheta + R^{-\alpha}{}_{\beta\mu\nu}(\1-\underline{\theta})+\underline{\delta}_\mu \left[\Gamma^\a_{\b\nu}\right] - \underline{\delta}_\nu \left[\Gamma^\a_{\b\mu}\right].\label{Riedist}
\ee
The last two terms here define the {\em singular} part of the Riemann tensor distribution, with support on $\S$, given and denoted by
\bea
\underline{H}^\a{}_{\b\mu\nu} \defi \underline{\delta}_\mu \left[\Gamma^\a_{\b\nu}\right] - \underline{\delta}_\nu \left[\Gamma^\a_{\b\mu}\right]=\underline{\delta}_\nu\left(n^\a y_{\b\mu}-n_\b y^\a_\mu \right)-\underline{\delta}_\mu\left(n^\a y_{\b\nu}-n_\b y^\a_\nu \right)\nonumber\\
=\deltaS \left\{n_\nu\left(n^\a y_{\b\mu}-n_\b y^\a_\mu \right)-n_\mu \left(n^\a y_{\b\nu}-n_\b y^\a_\nu \right) \right\}\defi \deltaS H^\a{}_{\b\mu\nu}\label{Hriemann}
\eea
where we have used (\ref{Gammajump}) and (\ref{delta}). Again, $H^\a{}_{\b\mu\nu}$ is a tensor field defined only on $\S$ independent of the choice of rigging but affected by the choice of normalization of the normal one-form $\bm{n}$, so that only its product with $\deltaS$, denoted by $\underline{H}^\a{}_{\b\mu\nu}$,  is univocally defined. Note that
\be
\underline{H}^\a{}_{\b[\mu\nu}n_{\tau]}=0 =H^\a{}_{\b[\mu\nu}n_{\tau]} .\label{Hn}
\ee
From $\underline{H}^\a{}_{\b\mu\nu}$ one can immediately obtain the singular parts of the Ricci tensor, scalar curvature, and Einstein tensor distributions, given respectively by
\bea
\underline{H}_{\b\nu}\defi  \underline{H}^\m{}_{\b\mu\nu} &=&\deltaS\left(  n^\rho y_{\b\rho} n_\nu+n^\rho y_{\nu\rho} n_\b -\nn y_{\b\nu} - y^\rho_\rho n_\b n_\nu\right)\defi \deltaS H_{\b\nu}\label{Hricci}\\
\underline{H} \defi  g^{\b\m}\underline{H}_{\b\m} &=&2\deltaS \left(n^\rho n^\sigma y_{\rho\sigma} -\nn y^\rho_\rho \right)\defi \deltaS H \label{Hscalar}\\
 \underline{{\cal G}}_{\b\mu} \defi \underline{H}_{\b\mu}- \frac{1}{2} \underline{H} g_{\b\mu}  &=&
\deltaS \left(  n^\rho y_{\b\rho} n_\nu+n^\rho y_{\nu\rho} n_\b -\nn y_{\b\nu} - y^\rho_\rho n_\b n_\nu\right.\nonumber \\
&&\left. -g_{\b\m}(n^\rho n^\sigma y_{\rho\sigma} -\nn y^\rho_\rho ) \right)\defi \deltaS {\cal G}_{\b\mu}\label{G}.
\eea
The latter formula is the generalization of the Israel equation to general shells, first found in \cite{MS} (see also \cite{MSV,mars}), and it reduces to the one found in \cite{BI} for null shells when $\nn =0$, and to the original Israel equation \cite{I} for non-null shells in the canonical gauge:
$$
{\cal G}_{\b\m} = -\left[K_{\b\m}\right] +\left(g_{\b\m}-\epsilon n_\b n_\m \right) \left[K^\rho_\rho\right]\hs \mbox{(non-null $\S$ only, $\nn =\epsilon$)}.
$$
In General Relativity, (\ref{G}) gives the singular part $\tau_{\m\n}\deltaS$ with support on $\S$ of $\underline{T}_{\m\n}$ via the Einstein field equations, see section \ref{Conclusions}. Nevertheless, I will generically keep the Einstein singular part ${\cal G}_{\m\n}$ in all the formulas because one can apply them, if desired, to other gravitational theories different from General Relativity, e.g. \cite{Senovilla13,RSV}. 

In general, it is easily checked that
$$
n^\b \underline{{\cal G}}_{\b\mu} =0= n^\b {\cal G}_{\b\mu} .
$$
Therefore (i) at points where $\S$ is non-null ${\cal G}_{\m\n}$ is tangent to $\S$ and (ii) at points where $\S$ is null ---and thus $\nn =0$--- ${\cal G}^{\m\n}$ has no non-zero component transversal to $\S$ there. Perhaps a better way, for general shells, to look at this identity is in the form
\be
n_\m {\cal G}^{\m\n}=0, \hs \Longrightarrow \hs {\cal G}^{\m\n}={\cal G}^{ab} e^\m_a e^\n_b \label{nG}
\ee
which implies that the contravariant version ${\cal G}^{\m\n}$ possesses only components tangential to $\S$ ---and this statement is independent of the choice of rigging.

Finally, I consider the relation between the curvature of the ambient manifold $(M,g)$ and the curvature of the general hypersurface $\S$ with the rigged connection. In contrast with the cases with non-null $\S$, for general hypersurfaces with null points one must distinguish three different Codazzi relations, together with the Gauss equation, see \cite{MS}. The relations are given by
\bea
\omega^d_\a R^{\pm\a}{}_{\b \g \d}e_a^\b e_b^\g e_c^\d = \Rb^{\pm d}{}_{abc} - K^\pm_{ac}\Psi_{\pm b}^d + K^\pm_{ab}\Psi^d_{\pm c}, \label{gauss}\\
n_\m R^{\pm\m}{}_{\a\b\g}e_a^\a e^\b_b e^\g_c = \overline{\N}^\pm_c K^\pm_{ba} - \overline{\N}^\pm_b K^\pm_{ca} +\varphi^\pm_c K^\pm_{ba} - \varphi^\pm_b K^\pm_{ca}, \label{codazzi1}\\
\omega^d_\a R^{\pm\a}{}_{\b \g \d}\ell^\b e_b^\g e_c^\d = \overline{\nabla}^\pm_b \Psi^d_{\pm c}- \overline{\nabla}^\pm_c \Psi^d_{\pm b}-\varphi^\pm_b \Psi^d_{\pm c}+\varphi^\pm_c \Psi^d_{\pm b},
\label{codazzi2}\\
n_\a R^{\pm\a}{}_{\b \g \d}\ell^\b e_b^\g e_c^\d = \overline{\nabla}^\pm_b\varphi^\pm_c-\overline{\nabla}^\pm_c \varphi^\pm_b+K^\pm_{ab}\Psi^a_{\pm c}-K^\pm_{ac}\Psi^a_{\pm b}.\label{codazzi3}
\eea
where $\Rb^{\pm d}{}_{abc}$ are the curvature tensors of the rigged connections $\bar{\Gamma}^{\pm a}_{bc}$

\end{document}